\documentclass[]{raa}            

\usepackage{graphicx,times}             
\usepackage{natbib}
\usepackage{amssymb,amsmath}
\bibpunct{(}{)}{;}{a}{}{,}
\usepackage[a4paper=true,dvipdfm=true,pagebackref=true]{hyperref}
\hypersetup{colorlinks = true, linkcolor = green, anchorcolor = red, citecolor = blue, filecolor = red, pagecolor = red, urlcolor = red}

\graphicspath{{img/}}
\usepackage{color}
\usepackage{ulem}

\usepackage[labelsep=space]{caption}
\usepackage{fancyhdr}

\fancypagestyle{cover}{%
	
	\lhead{}
	\rhead{Apr. 9, 2018}
	\cfoot{}}

\begin{document}
\title{Photometric investigations on two totally eclipsing contact binaries: V342 UMa and V509 Cam
}

   \volnopage{Vol.0 (20xx) No.0, 000--000}      
   \setcounter{page}{1}          

   \author{Li Kai\inst{1,2}
   \and Xia, Qi-Qi\inst{1}
   \and Liu, Jin-Zhong\inst{3}
   \and Zhang, Yu\inst{3}
   \and Gao, Xing\inst{3}
   \and Hu, Shao-Ming\inst{1}
   \and Guo, Di-Fu\inst{1}
   \and Chen, Xu\inst{1}
   \and Liu, Yuan\inst{4}
   }

   \institute{Shandong Provincial Key Laboratory of Optical Astronomy and Solar-Terrestrial Environment, Institute of Space Sciences, Shandong University, Weihai, 264209, China; {\it kaili@sdu.edu.cn (Li, Kai)}\\
        \and
             Key Laboratory for the Structure and Evolution of Celestial Objects, Chinese Academy of Sciences\\
        \and
             Xinjiang Astronomical Observatory, Chinese Academy of Sciences, Urumqi, 830011, China\\
        \and
             Qilu Institute of Technology, Jinan, 250200, China\\
\vs\no
   {\small Received~~20xx month day; accepted~~20xx~~month day}}

\abstract{ By analyzing two sets of complete $BVR_cI_c$ light curves of V342 UMa and three sets of complete $BVR_cI_c$ light curves of V509 Cam, we determined that the two systems are both W-subtype contact binaries and that V342 UMa shows a shallow contact configuration, while V509 Cam exhibits a medium contact configuration. Since both of them are totally eclipsing binaries, the physical parameters derived only by the photometric light curves are reliable.Meanwhile, the period changes of the two targets were analyzed based on all available eclipsing times. We discovered that V342 UMa shows long-term period decrease with a rate of $-1.02(\pm0.54)\times10^{-7}$ days/year and that V509 Cam exhibits long-term period increase with a rate of $3.96(\pm0.90)\times10^{-8}$ days/year. Both the conservative mass transfer and AML via magnetic stellar winds can interpret the long-term period decrease of V342 UMa. The long-term period increase of V509 Cam can be explained by mass transfer from the less massive star to the more massive one. The absolute parameters of the two binaries were estimated according to the Gaia distances and our derived photometric solution results. This method can be extended to other contact binaries without radial velocities but with reliable photometric solutions. The evolutionary states of them were discussed, we found that they reveal identical properties of other W-subtype contact systems.
}

   \authorrunning{Li et al.}            
   \titlerunning{Two totally eclipsing contact binaries}  

   \maketitle

%
%
\section{Introduction}           
\label{sect:intro}

W UMa contact binaries are comprised of two late type stars with spectral types from F to K. The two component stars are sharing a common convective envelope and have nearly equal effective temperatures although their masses are very different. The analysis of W UMa contact binaries is very necessary for modern astrophysics as they are probes for understanding tidal interactions, energy exchange, mass transfer, and angular momentum loss. The formation, evolution, ultimate fate, and magnetic activities of the W UMa contact binaries are still debatable issues (e.g., \citealt{Guinan+Bradstreet+1988, Bradstreet+Guinan+1994, Eggleton+Kisseleva-Eggleton+2006, Fabrycky+Tremaine+2007, Qian+etal+2006, Qian+etal+2007a, Qian+etal+2014, Qian+etal+2017, Qian+etal+2018}). In order to solve these problems, the determination of physical parameters of a great deal of such type binaries is required.

The physical parameters, such as mass ratio, are poorly estimated for partially eclipsing contact binaries (e.g., \citealt{Pribulla+etal+2003, Terrell+Wilson+2005}). In addition, the spectroscopic mass ratio sometimes can not be reliably derived according to the broadened and blended spectral lines (e.g., \citealt{Dall+Schmidtobreick+2005, Rucinski+2010}). By the study of contact binaries that have been obtained both spectroscopic and photometric mass ratios, \cite{Pribulla+etal+2003} discovered that the photometric mass ratios of the totally eclipsing systems correspond to their spectroscopic ones. \cite{Terrell+Wilson+2005} determined similar result by discussing the relations between photometric and spectroscopic mass ratios. These results suggest that we can derive very precise and reliable physical parameters for totally eclipsing contact binaries only by the photometric light curves. Thanks to the Gaia mission (\citealt{Gaia Collaboration+etal+2018}), the parallaxes of more than one billion stars have been obtained, which allows researchers to estimate the absolute parameters of contact binaries even if there is no radial velocity observations (e.g., \citealt{Kjurkchieva+etal+2019a, Kjurkchieva+etal+2019b}). Therefore, we chose two totally eclipsing binaries, V342 UMa and V509 Cam, to analyze their light curves and period variations and estimate their absolute parameters.

V342 UMa was firstly discovered as a W UMa type binary by \cite{Nelson+etal+2004} during the observations of a nearby star BH UMa. The period of 0.343854 days, the color index of $B-V=0.64$, and spectral type of G3 were obtained. A photometric study by them revealed that V342 UMa is a low mass ratio ($q=0.331$) W-subtype contact binary (the hotter component is the less massive one). It has been fifteen years after the discovery and the first photometric investigation of V342 UMa, we decided to investigate the light curves and orbital period changes of this target.

V509 Cam was firstly identified as an EW type eclipsing binary by \cite{Khruslov+2006} during an eclipsing binaries search in Camelopardalis. The variability amplitude of 0.6 mag and the orbital period of 0.35034 days were determined by him. At present, neither the light curve synthesis nor period variation analysis has been carried out for this star, we will analyze the light curves and orbital period variations of this target in this paper.

\section{CCD observations of V342 UMa and V509 Cam}
\label{sect:Obs}

\begin{figure}
\centering
\includegraphics[width=0.5\textwidth]{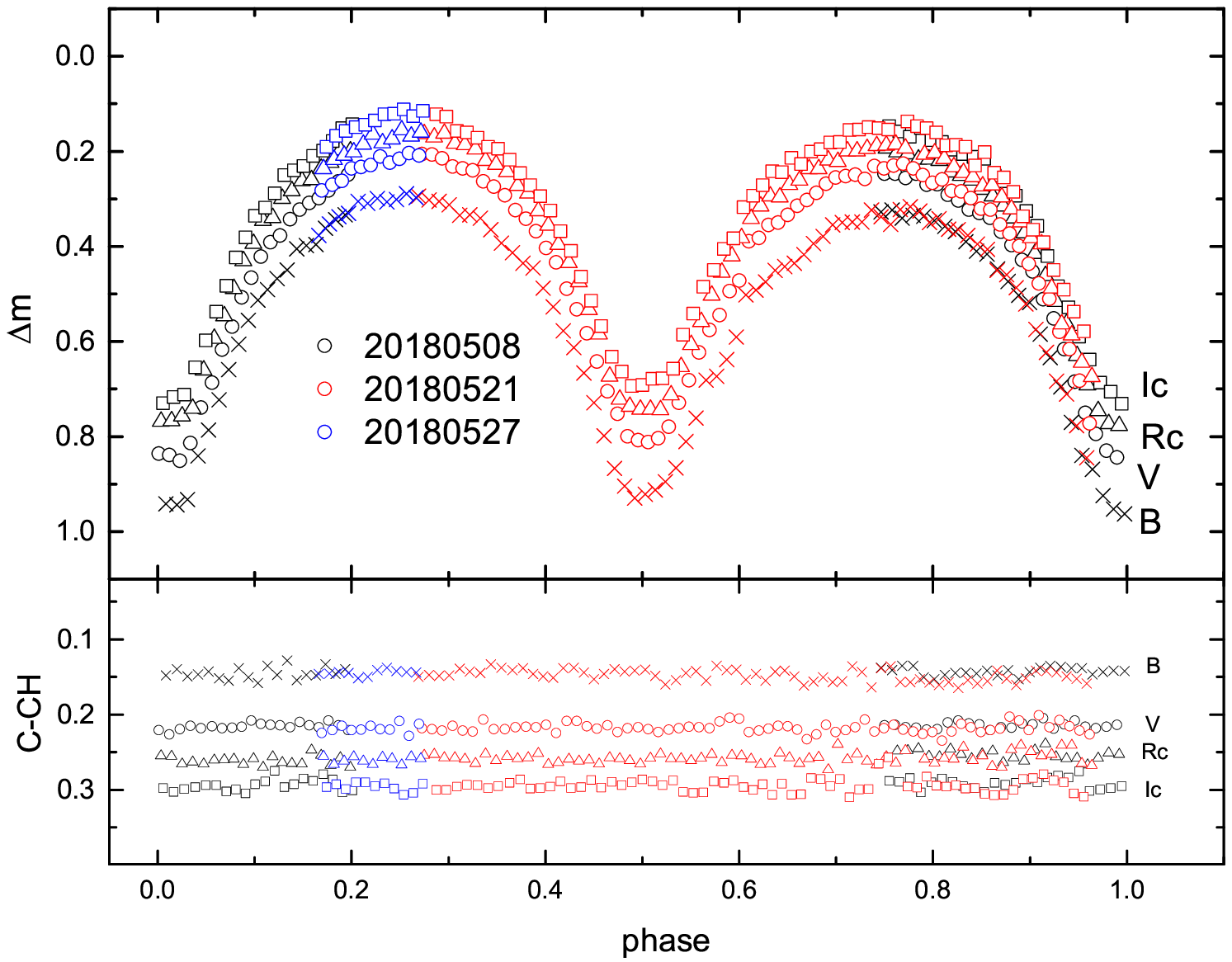}%
\includegraphics[width=0.5\textwidth]{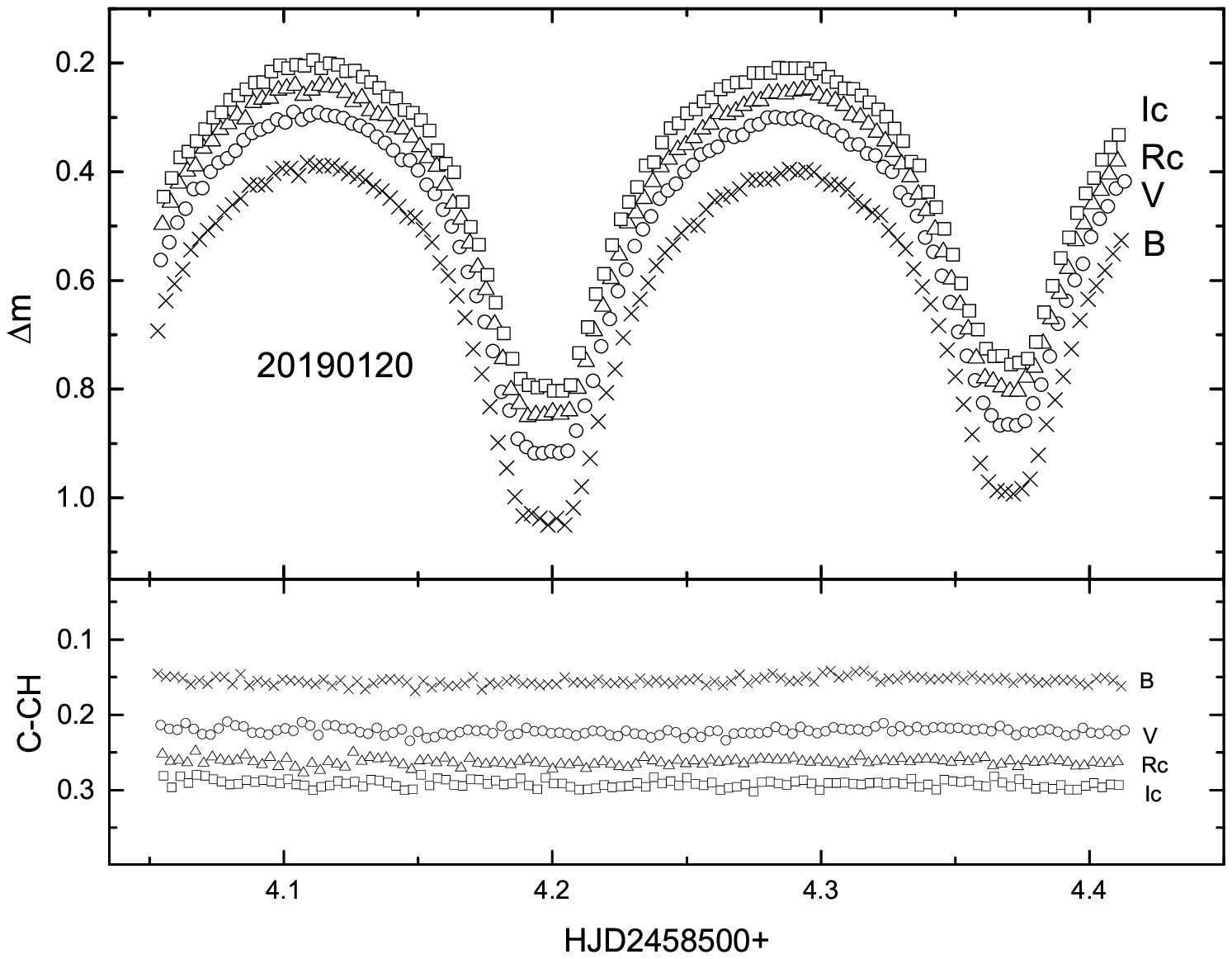}

\caption{The left figure displays the light curves of V342 UMa observed by NEXT on May 08, 21, and 27, 2018 (black, red and blue symbols respectively represent 20180508 observations, 20180521 observations, and 20180527 observations), while the right figure displays the light curves of V342 UMa observed by WHOT on Jan 20, 2019.
Crosses refer to the $B$ band light curves, while open circles, triangles, and squares respectively represent the $V$, $R_c$, and $I_c$ bands light curves. }
\label{Fig1}
\end{figure}

\begin{figure}
\centering
\includegraphics[width=0.33\textwidth]{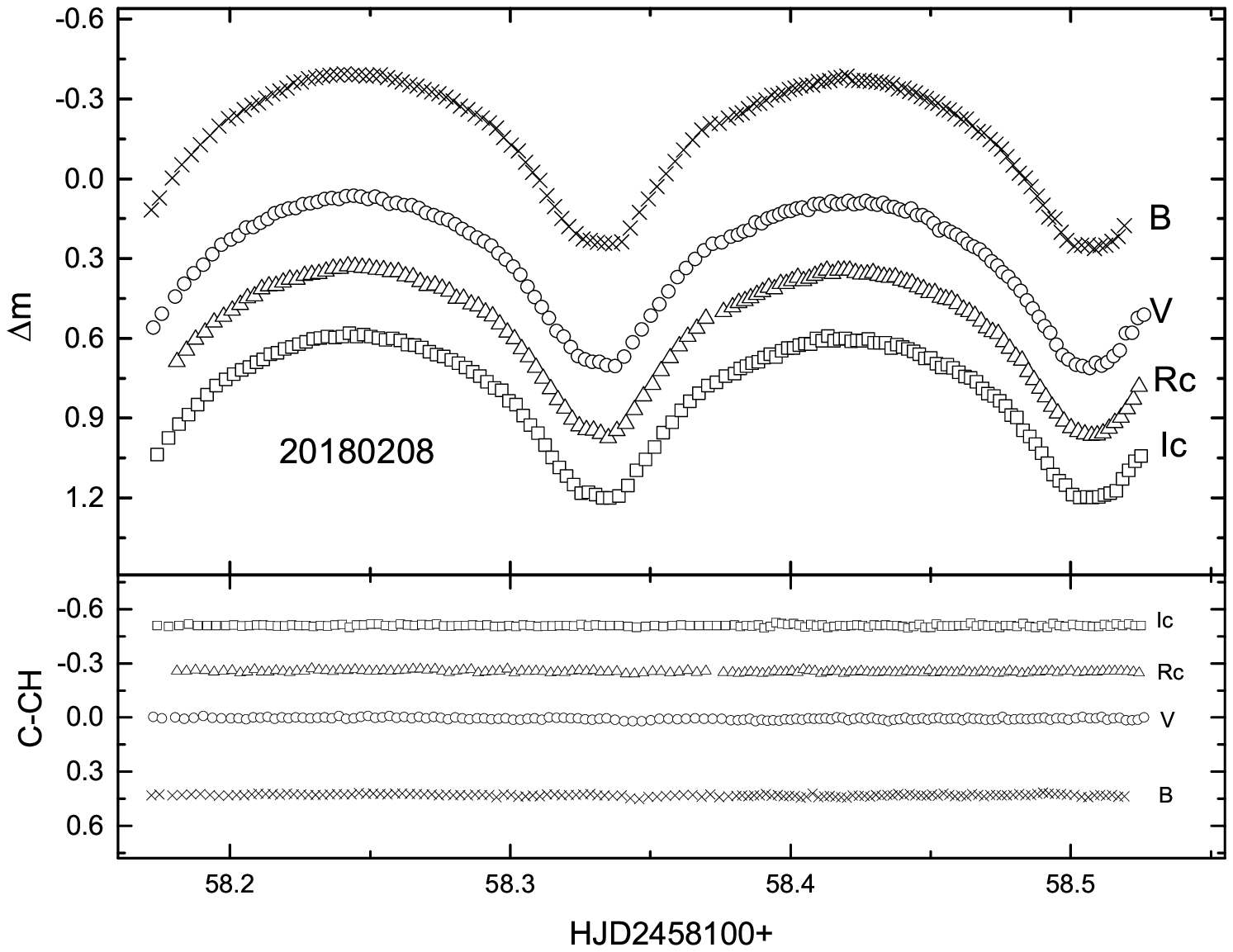}%
\includegraphics[width=0.33\textwidth]{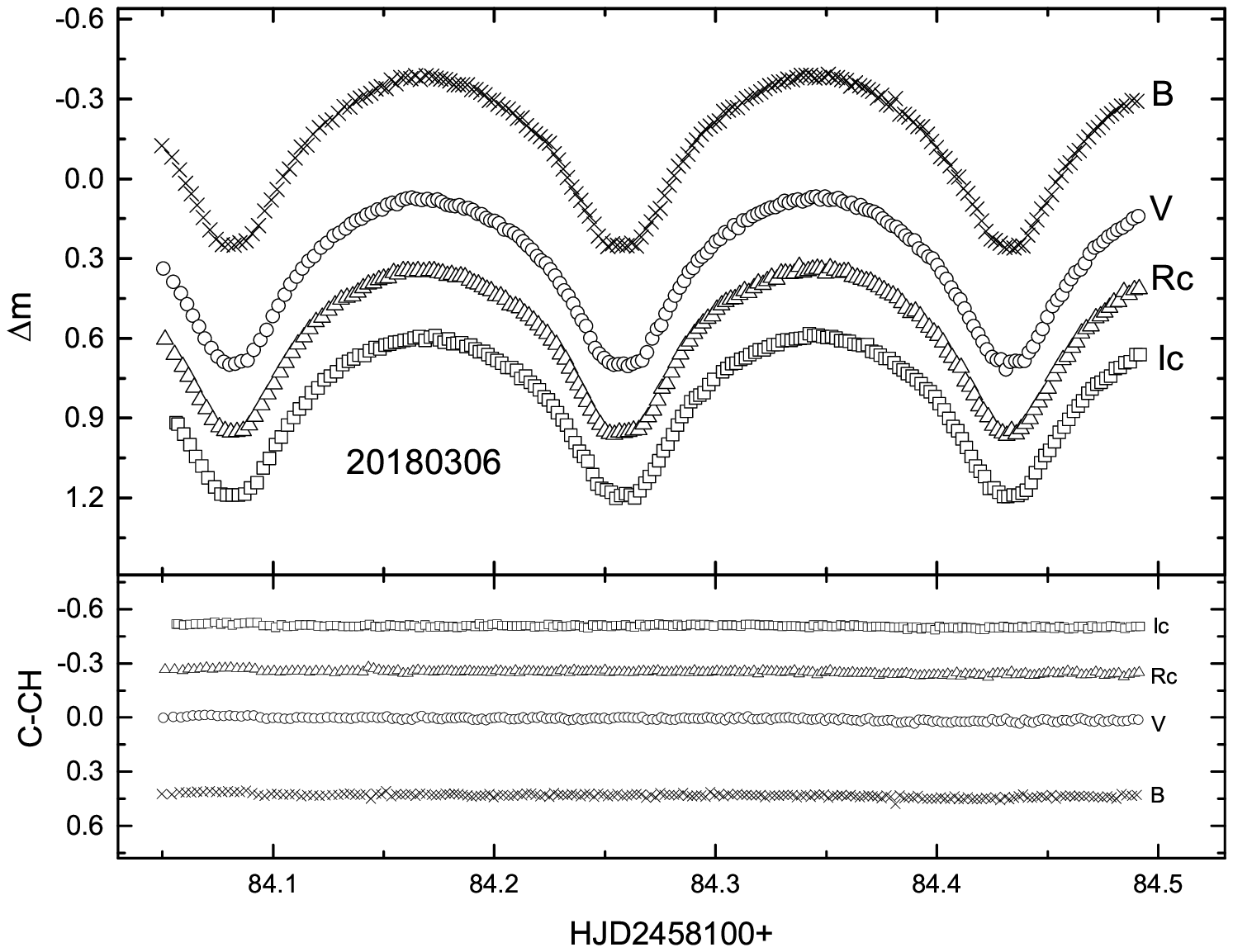}
\includegraphics[width=0.33\textwidth]{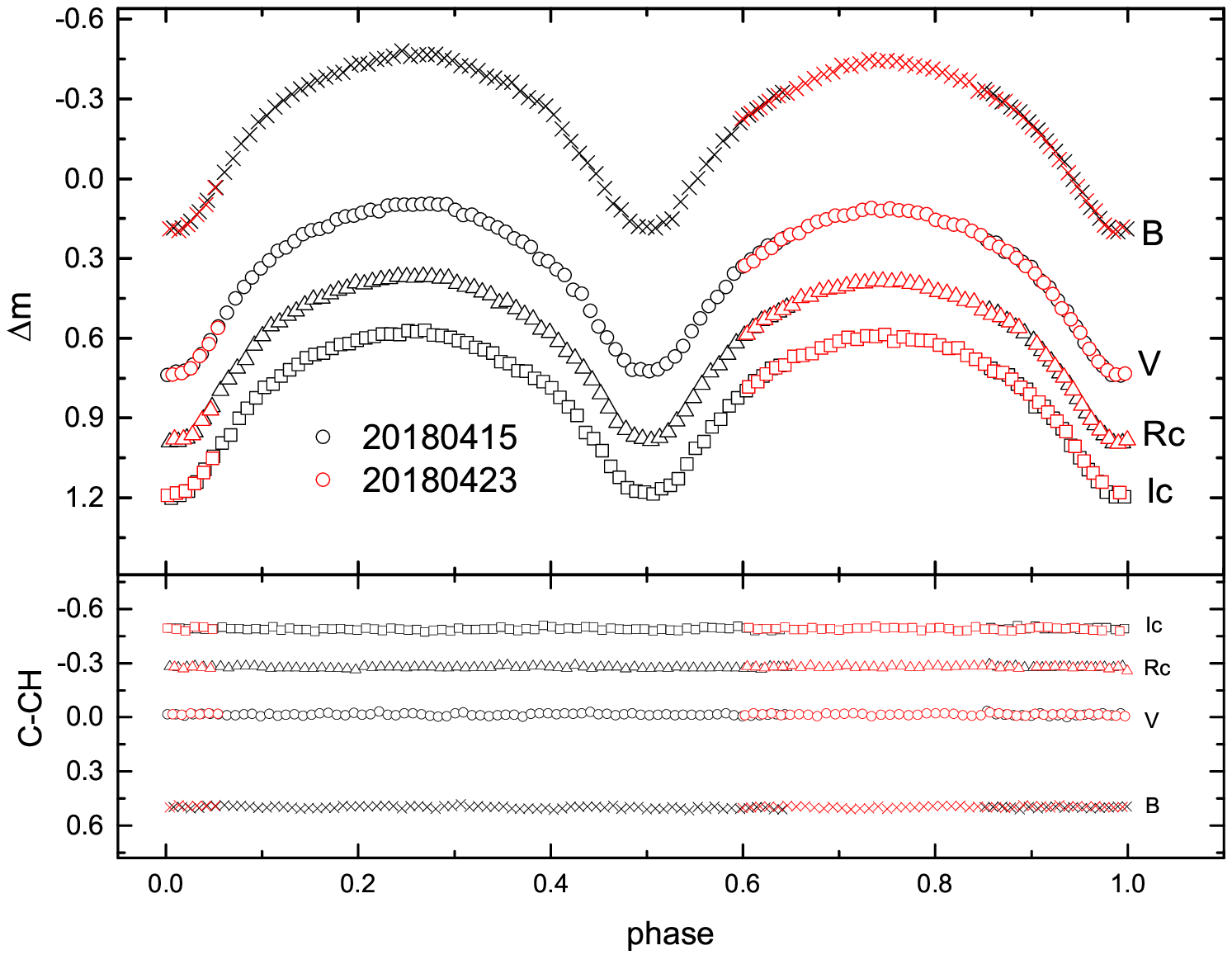}

\caption{The left panel displays the light curves of V509 Cam observed by NOWT on February 08, 2018, the middle panel shows the light curves of V509 Cam observed by NOWT on Mar 06, 2018, while the right panel plots the light curves of V509 Cam observed by NEXT on April 15 and 23, 2018 (black and red symbols respectively represent 20180415 observations and 20180423 observations).
Crosses refer to the $B$ band light curves, while open circles, triangles, and squares respectively represent the $V$, $R_c$, and $I_c$ bands light curves. }
\label{Fig2}
\end{figure}

Charge-coupled device (CCD) photometry of V342 UMa and V509 Cam were carried out from 2018 to 2019 using the Weihai Observatory 1.0-m telescope of Shandong University (WHOT, \citealt{Hu+etal+2014}), the Nanshan One-meter Widefield Telescope (NOWT, \citealt{Liu+etal+2014}) at the Nanshan station of the Xinjiang Astronomical Observatory, the 60cm Ningbo Bureau of Education and Xinjiang Observatory Telescope (NEXT), and the 85 cm telescope at the Xinglong Station of National Astronomical Observatories (NAOs85cm) in China. The observational information is listed in Table~\ref{Tab:obsevation}. In order to record the observed images, $2K\times2K$ CCD cameras were used for WHOT, NEXT, and NAOs85cm, and a $4K\times4K$ CCD camera was applied to NOWT. The field of views are $12^{'}\times12^{'}$ for WHOT, $1.3^{\circ}\times1.3^{\circ}$ for NOWT, $22^{'}\times22^{'}$ for NEXT, and $32^{'}\times32^{'}$ for NAOs85cm. The effective subframe of NOWT is $30^{'}\times30^{'}$ during the observations. The filters we used are standard Johnson-Cousin-Bessel $BVR_cI_c$ systems. The standard IRAF routine was applied to process the observed data including zero and flat calibrations, and aperture photometry, then different magnitudes between the target and the comparison star and those between the comparison and check stars were obtained. The complete light curves of V342 UMa observed by NEXT and WHOT and those of V509 Cam observed by NOWT and NEXT are illustrated in Figure~\ref{Fig1} and Figure~\ref{Fig2}, respectively. As seen in the two figures, the two targets show EW type light curves, and very clearly flat primary minima can be discovered. Based on our observations, six eclipsing minima were derived for V342 UMa, while ten were obtained for V509 Cam, all the minima were calculated by the K-W method (\citealt{Kwee+van Woerden+1956}) and are listed in Table~\ref{Tab:ecl-times}.

\begin{table}
\footnotesize
\begin{center}
\caption[]{The Observational Log for V342 UMa and V509 Cam.}
  \label{Tab:obsevation}
\begin{tabular}{cccccc}
\hline\hline
Star&Date	      & Filters and Typical exposure time	       & Type           &Uncertainties (mag)$^*$ & Telescope             \\ \hline
V342 UMa&Mar 29, 2018	&  $R_c$40s	                         & minimum light  &$R_c$0.011 &WHOT      \\
& May 08, 2018	& $B$80s $V$60s $R_c$40s $I_c$40s          & light curve    &$B$0.007 $V$0.006 $R_c$0.007 $I_c$0.009 &NEXT\\
& May 21, 2018	& $B$80s $V$60s $R_c$40s $I_c$40s 	       & light curve    &$B$0.010 $V$0.007 $R_c$0.007 $I_c$0.010 &NEXT       \\
& May 27, 2018	& $B$80s $V$60s $R_c$40s $I_c$40s  	       & light curve    &$B$0.008 $V$0.006 $R_c$0.007 $I_c$0.008 &NEXT \\
&Dec 28, 2018	&  $R_c$30s	                                 & minimum light  &$R_c$0.005 &NAOs85cm      \\
&Jan 20, 2019	&  $B$120s $V$60s $R_c$35s $I_c$25s	         & light curve    &$B$0.005 $V$0.005 $R_c$0.005 $I_c$0.005 &WHOT      \\
V509 Cam& Feb 08, 2018	& $B$36s $V$22s $R_c$13s $I_c$12s	 & light curve    &$B$0.006 $V$0.006 $R_c$0.006 $I_c$0.005 & NOWT     \\
& Mar 05, 2018	& $B$25s $V$25s $R_c$25s $I_c$25s	         & minimum light  &$B$0.010 $V$0.009 $R_c$0.008 $I_c$0.008 & NOWT     \\
& Mar 06, 2018	& $B$14s $V$10s $R_c$10s $I_c$10s	         & light curve    &$B$0.006 $V$0.005 $R_c$0.005 $I_c$0.005 & NOWT     \\
& Apr 15, 2018	& $B$70s $V$50s $R_c$40s $I_c$30s  	       & light curve    &$B$0.007 $V$0.006 $R_c$0.005 $I_c$0.007 &NEXT \\
& Apr 23, 2018	& $B$70s $V$50s $R_c$40s $I_c$30s  	       & light curve    &$B$0.006 $V$0.005 $R_c$0.005 $I_c$0.007 &NEXT \\
&Jan 21, 2019	&  $R_c$40s 	                               & minimum light  &$R_c$0.003 &WHOT      \\
\hline
\end{tabular}
\end{center}
$^*$ The uncertainties are the standard deviation of the differences between the comparison and check stars.
\end{table}

\section{Orbital period variations}
\label{sect:orbtial}

\begin{figure}
\centering
\includegraphics[width=0.5\textwidth]{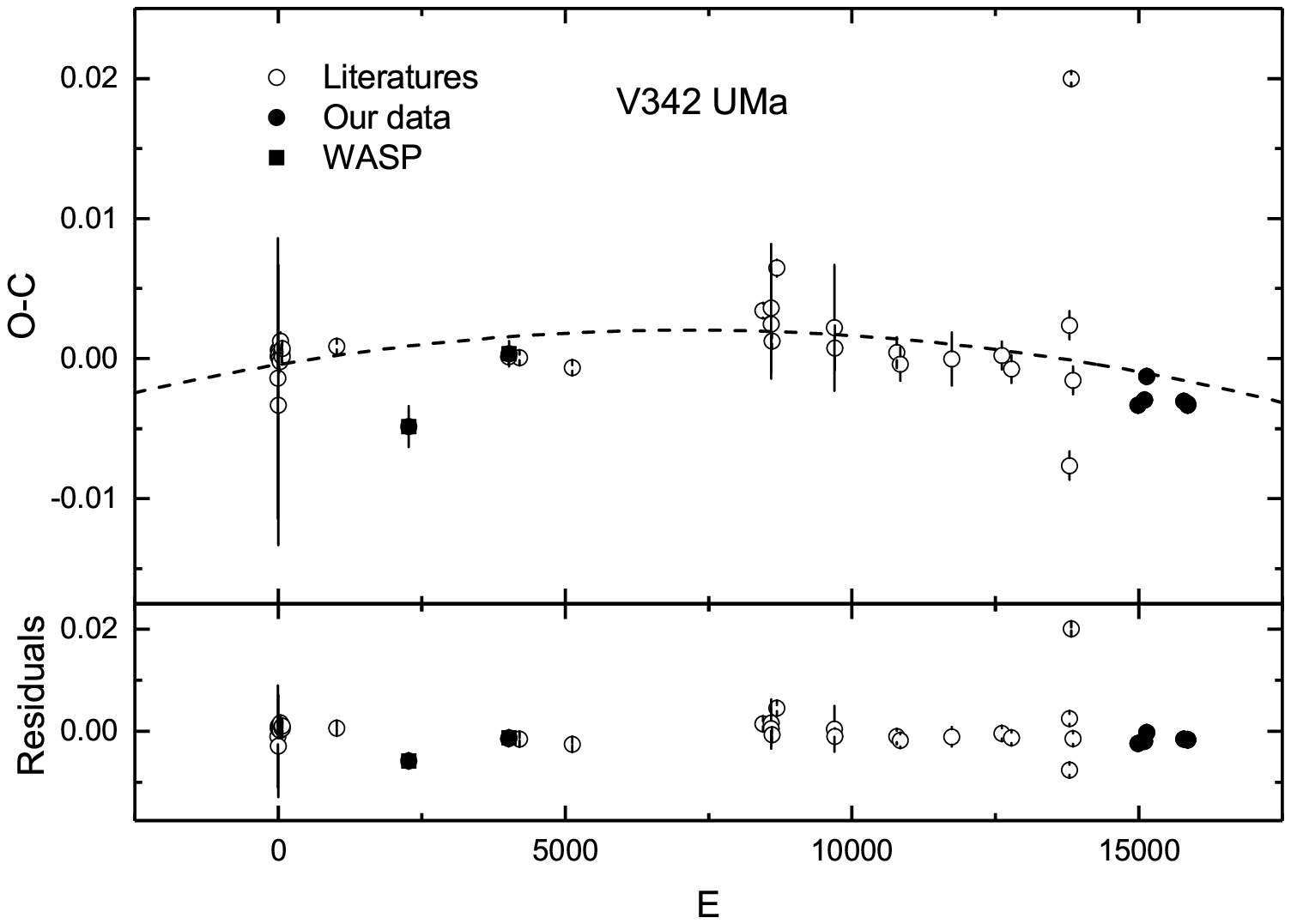}%
\includegraphics[width=0.505\textwidth]{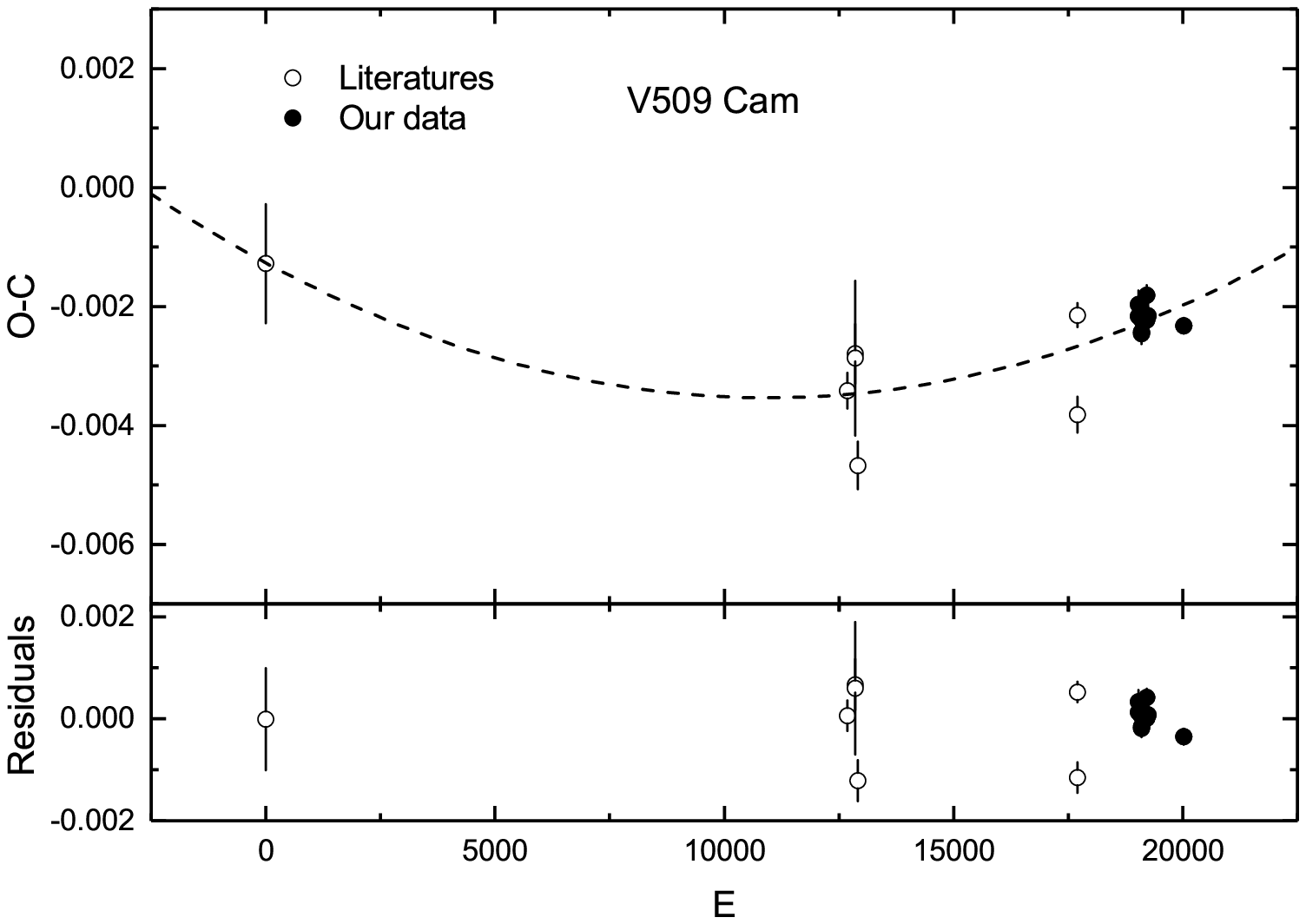}

\caption{The left panel displays the O-C curve of V342 UMa, while the right panel refer to the O-C curve of V509 Cam.
Open circles refer to eclipsing times from literatures, solid circles represent our data, while solid squares show the WASP eclipsing times. The errors of some points were not given in the literature and are fixed at 0.0010 when plotting this figure.}
\label{Fig3}
\end{figure}

Both of V342 UMa and V509 Cam have been identified more than ten years, no one has analyzed the orbital period variations at present. Then, we collected all published eclipsing times for V342 UMa and V509 Cam from literatures, and listed them in Table~\ref{Tab:ecl-times}. Moreover, WASP (Wide Angle Search for Planets, \citealt{Butters+etal+2010}) has observed V342 UMa, we calculated two minima using the archive data and also listed in Table~\ref{Tab:ecl-times}. Combining our new observed ones, we obtained a total of 43 photoelectric or CCD eclipsing times for V342 UMa, and a total of 17 photoelectric or CCD eclipsing times for V509 Cam. Using the least-squares method, the linear ephemeris of V342 UMa taken from \cite{Nelson+etal+2004} was corrected to be:
\begin{equation}
\textrm{Min.I}=2453054.83723(\pm0.00090)+0.34385184(\pm0.00000010)\textrm{E},
\end{equation}
and the linear ephemeris of V509 Cam originated from O-C Gateway\footnote{The website of O-C Gateway is http://var.astro.cz/ocgate/} was amended to be:
\begin{equation}
\textrm{Min.I}=2451492.27572(\pm0.00071)+0.35034717(\pm0.00000004)\textrm{E}.
\end{equation}
All the O-C values calculated by the two equations are listed in Table~\ref{Tab:ecl-times}, the corresponding curves are displayed in Figure~\ref{Fig3}. We can find that both of V342 UMa and V509 Cam show a parabolic trend. Then, quadratic ephemerides were applied to fit the O-C curves of the two targets,
\begin{equation}
\textrm{Min.I}=-0.00043(\pm0.00095)+0.00000069(\pm0.00000038)\textrm{E}-4.81(\pm2.56)\times10^{-11}\textrm{E}^2 ,
\end{equation}
\begin{equation}
\textrm{Min.I}=-0.00127(\pm0.00056)-0.00000041(\pm0.00000010)\textrm{E}+1.90(\pm0.43)\times10^{-11}\textrm{E}^2 .
\end{equation}
When removing Equations (3) and (4), the residuals are listed in Table~\ref{Tab:ecl-times} and shown in the bottom panels of Figure~\ref{Fig3}. No cyclic variations can be detected in the residuals. According to the coefficients of the quadratic terms of Equations (3) and (4), we determined that the period of V342 UMa is secular decrease at a rate of $-1.02(\pm0.54)\times10^{-7}$ days/year and the period of V509 Cam is continuously increase at a rate of $3.96(\pm0.90)\times10^{-8}$ days/year.

\begin{table}
\tiny
\begin{center}
\caption{Eclipsing Times of V342 UMa and V509 Cam}
\label{Tab:ecl-times}
\begin{tabular}{cccccc|cccccc}
\hline
              \multicolumn{ 6}{c}{V342 UMa} &        \multicolumn{ 6}{c}{ V509 Cam}       \\\hline
HJD         &  Errors &  E        & O-C     &  Residuals& References& HJD        & Errors & E       & O-C     & Residuals& References\\
2400000+    &         &           &         &           &           & 2400000+   &        &         &         &          &           \\\hline
53053.8040  &  0.0100 &  -3       & -0.0014 &  -0.0010  & (1) & 51492.2770 & $-$    & 0       & -0.0013 & 0.0000   & (15)       \\
53053.9740  &  0.0100 &  -2.5     & -0.0033 &  -0.0029  & (1) & 55937.8301 & 0.0003 & 12689   & -0.0034 & 0.0001   & (6) \\
53054.8371  &  0.0001 &  0        & 0.0001  &  0.0006   & (1) & 55996.3387 & 0.0005 & 12856   & -0.0028 & 0.0007   & (7) \\
53055.3529  &  0.0003 &  1.5      & 0.0002  &  0.0006   & (1) & 55996.5138 & 0.0013 & 12856.5 & -0.0029 & 0.0006   & (7) \\
53055.5253  &  0.0002 &  2        & 0.0006  &  0.0011   & (1) & 56017.7080 & 0.0004 & 12917   & -0.0047 & -0.0012  & (6) \\
53057.7602  &  0.0002 &  8.5      & 0.0005  &  0.0009   & (1) & 57692.3700 & 0.0002 & 17697   & -0.0021 & 0.0005   & (16) \\
53057.9317  &  0.0002 &  9        & 0.0001  &  0.0005   & (1) & 57692.5435 & 0.0003 & 17697.5 & -0.0038 & -0.0012  & (16) \\
53058.9635  &  0.0002 &  12       & 0.0003  &  0.0007   & (1) & 58158.3319 & 0.0002 & 19027   & -0.0020 & 0.0003   & (14)\\
53059.6509  &  0.0010 &  14       & 0.0000  &  0.0004   & (1) & 58158.5069 & 0.0001 & 19027.5 & -0.0022 & 0.0001   & (14)\\
53066.8716  &  0.0006 &  35       & -0.0002 &  0.0002   & (1) & 58183.3815 & 0.0002 & 19098.5 & -0.0022 & 0.0000   & (14)\\
53067.0450  &  0.0006 &  35.5     & 0.0013  &  0.0017   & (1) & 58184.0819 & 0.0002 & 19100.5 & -0.0025 & -0.0002  & (14)\\
53074.7807  &  0.0003 &  58       & 0.0003  &  0.0007   & (1) & 58184.2575 & 0.0002 & 19101   & -0.0021 & 0.0002   & (14)\\
53074.9528  &  0.0003 &  58.5     & 0.0005  &  0.0009   & (1) & 58184.4323 & 0.0002 & 19101.5 & -0.0024 & -0.0002  & (14)\\
53077.7032  &  0.0003 &  66.5     & 0.0001  &  0.0005   & (1) & 58224.1969 & 0.0001 & 19215   & -0.0022 & 0.0000   & (14)\\
53077.8758  &  0.0002 &  67       & 0.0007  &  0.0011   & (1) & 58224.3725 & 0.0002 & 19215.5 & -0.0018 & 0.0004   & (14)\\
53404.8790  &  0.0002 &  1018     & 0.0009  &  0.0006   & (2) & 58232.2550 & 0.0002 & 19238   & -0.0022 & 0.0001   & (14)\\
53837.4389  &  0.0015 &  2276     & -0.0049 &  -0.0015  & (3) & 58505.3504 & 0.0001 & 20017.5 & -0.0023 & -0.0004  & (14)\\
54435.9180  &  0.0001 &  4016.5   & 0.0001  &  -0.0025  & (4) &                                                                \\
54438.6690  &  0.0009 &  4024.5   & 0.0003  &  0.0015   & (3) &                                                                \\
54499.8744  &  0.0002 &  4202.5   & 0.0001  &  0.0017   & (5) &                                                                \\
54815.8735  &  0.0002 &  5121.5   & -0.0007 &  0.0005   & (5) &                                                                \\
55958.8411  &  0.0005 &  8445.5   & 0.0034  &  -0.0007  & (6) &                                                                \\
56009.3875  &  0.0046 &  8592.5   & 0.0036  &  0.0046   & (7) &                                                                \\
56009.5583  &  0.0039 &  8593     & 0.0025  &  0.0005   & (7) &                                                                \\
56013.6833  &  0.0002 &  8605     & 0.0012  &  -0.0010  & (6) &                                                                \\
56042.7440  &  0.0006 &  8689.5   & 0.0065  &  -0.0010  & (6) &                                                                \\
56387.4512  &  0.0045 &  9692     & 0.0022  &  -0.0018  & (8) &                                                                \\
56390.3725  &  0.0016 &  9700.5   & 0.0008  &  -0.0011  & (8) &                                                                \\
56761.3883  &  0.0011 &  10779.5  & 0.0004  &  0.0201   & (9) &                                                                \\
56783.3940  &  0.0012 &  10843.5  & -0.0004 &  -0.0014  & (9) &                                                                \\
57091.4856  &  0.0019 &  11739.5  & 0.0000  &  -0.0004  & (10) &                                                                \\
57390.2931  &  $-$    &  12608.5  & 0.0002  &  -0.0012  & (11)  &                                                                \\
57450.2943  &  $-$    &  12783    & -0.0007 &  0.0025   & (11)  &                                                                \\
57797.0720  &  $-$    &  13791.5  & 0.0024  &  -0.0076  & (12)  &                                                                \\
57797.2339  &  $-$    &  13792    & -0.0076 &  -0.0014  & (12)  &                                                                \\
57806.8894  &  0.0003 &  13820    & 0.0200  &  -0.0024  & (13) &                                                                \\
57817.1834  &  $-$    &  13850    & -0.0016 &  -0.0019  & (12)  &                                                                \\
58207.1096  &  0.0003 &  14984    & -0.0033 &  -0.0002  & (14)&                                                                \\
58247.3406  &  0.0003 &  15101    & -0.0030 &  -0.0015  & (14)&                                                                \\
58260.2368  &  0.0004 &  15138.5  & -0.0013 &  -0.0017  & (14)&                                                                \\
58481.3317  &  0.0001 &  15781.5  & -0.0030 &  -0.0016  & (14)&                                                                \\
58504.1976  &  0.0002 &  15848    & -0.0033 &  -0.0058  & (14)&                                                                \\
58504.3696  &  0.0002 &  15848.5  & -0.0032 &  -0.0012  & (14)&                                                                \\
\hline
\end{tabular}
\end{center}
\scriptsize
(1) \citealt{Nelson+etal+2004}; (2) \citealt{Krajci+2006}; (3) This paper (WASP); (4) \citealt{Nelson+2008}; (5) \citealt{Nelson+2009}; (6) \citealt{Diethelm+2012}; (7) \citealt{Hubscher+etal+2013}; (8) \citealt{Hubscher+2014}; (9) \citealt{Hubscher+Lehmann+2015}; (10) \citealt{Hubscher+2016}; (11) VSOLJ 63; (12) VSOLJ 64; (13) \citealt{Nelson+2018};
(14)  This paper (WHOT); (15) \citealt{Khruslov+2006}; (16) OEJV 0179. \\
\end{table}

\section{Photometric solutions of V342 UMa and V509 Cam}
\label{sect:photo}

Based on our observations, two sets of complete light curves of V342 UMa and three sets of light curves of V509 Cam were obtained. The Wilson-Devinney (W-D) program (\citealt{Wilson+Devinney+1971, Wilson+1979, Wilson+1990}) was used to model these light curves. Gaia DR 2 (\citealt{Gaia Collaboration+etal+2016, Gaia Collaboration+etal+2018}) has observed these two targets and determined the mean temperatures of them, $T_m=5741$ K for V342 UMa and $T_m=6462$ K for V509 Cam. At first, the mean temperature was set as the temperature of the primary, $T_1$. The bolometric and bandpass limb-darkening coefficients were taken from \cite{van Hamme+1993}'s table, and the gravity-darkening coefficients and the bolometric albedos were set as $g_{1,2}=0.32$ and $A_{1,2} = 0.5$ for their convective envelopes (\citealt{Lucy+1967, Rucinski+1969}). Due to the lack of radial velocity curves, the $q$-search method was applied to determine the mass ratios of the two systems. When we obtained the final solutions, the temperatures of the two components were calculated using the following method (\citealt{Coughlin+etal+2011, Dimitrov+Kjurkchieva+2015}),
\begin{eqnarray}
T_1&=&T_m+{c\Delta T\over c+1},  \\\nonumber
T_2&=&T_1-\Delta T,
\end{eqnarray}
where $\Delta T=T_1-T_2$ and $c=L_2/L_1$ are derived by the W-D modeling.

\begin{figure}
\begin{center}
\includegraphics[width=0.5\textwidth]{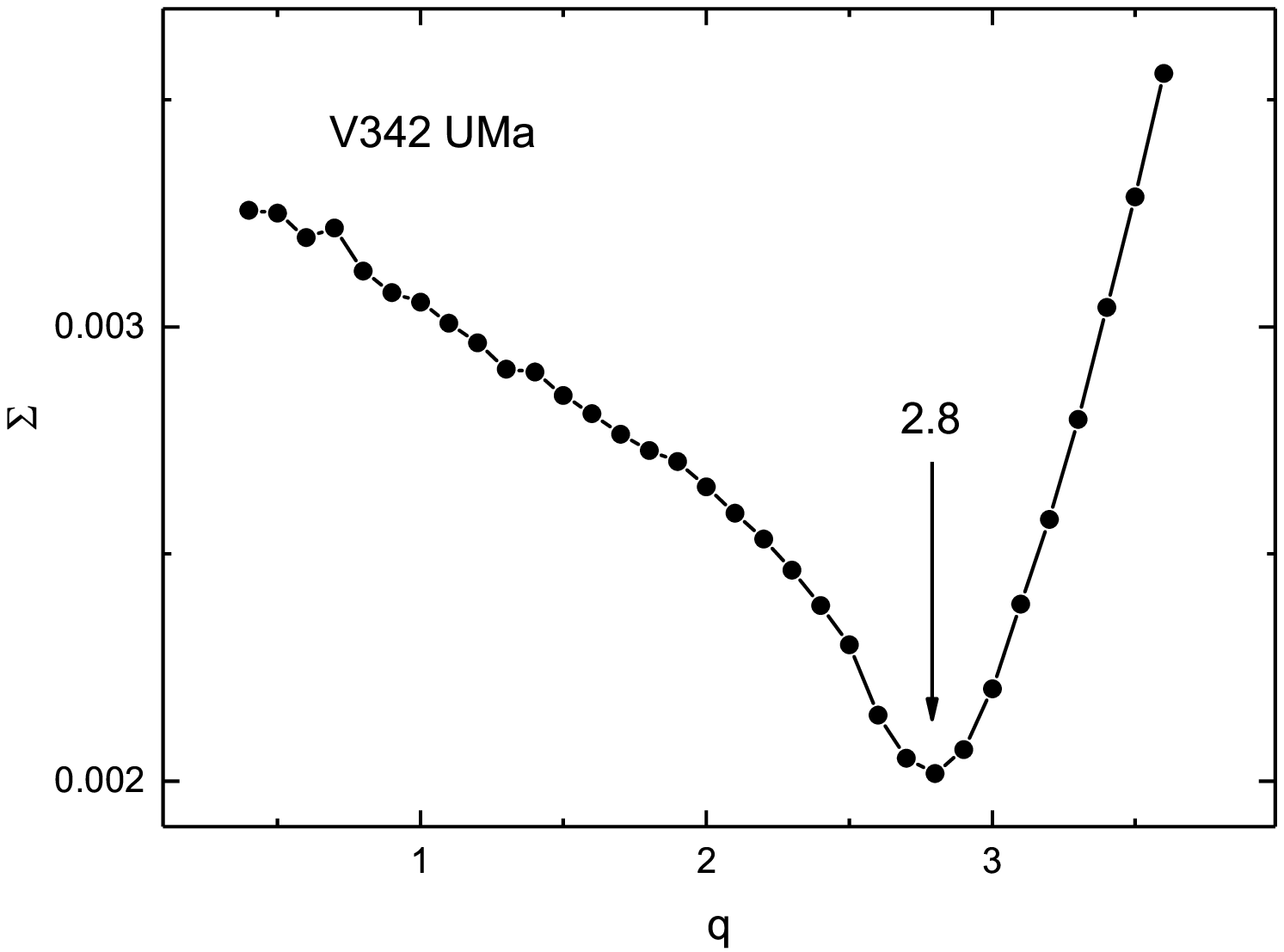}%
\includegraphics[width=0.5\textwidth]{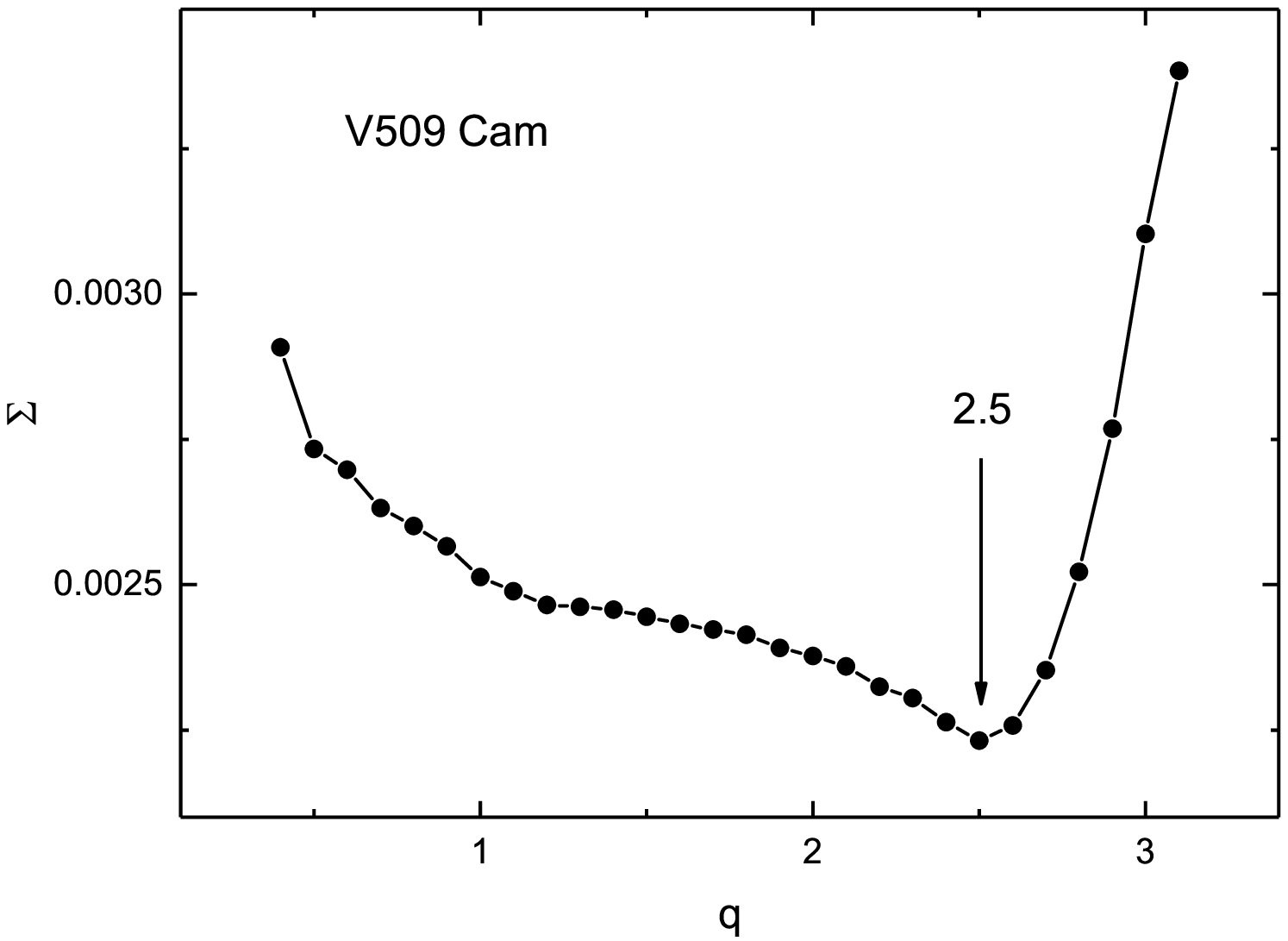}

\caption{ This figure displays the $\sum W_i(O-C)_i^2$ versus mass ratio $q$ of V342 UMa and V509 Cam. }
\label{Fig4}
\end{center}
\end{figure}

\begin{figure}
\begin{center}
\includegraphics[width=0.5\textwidth]{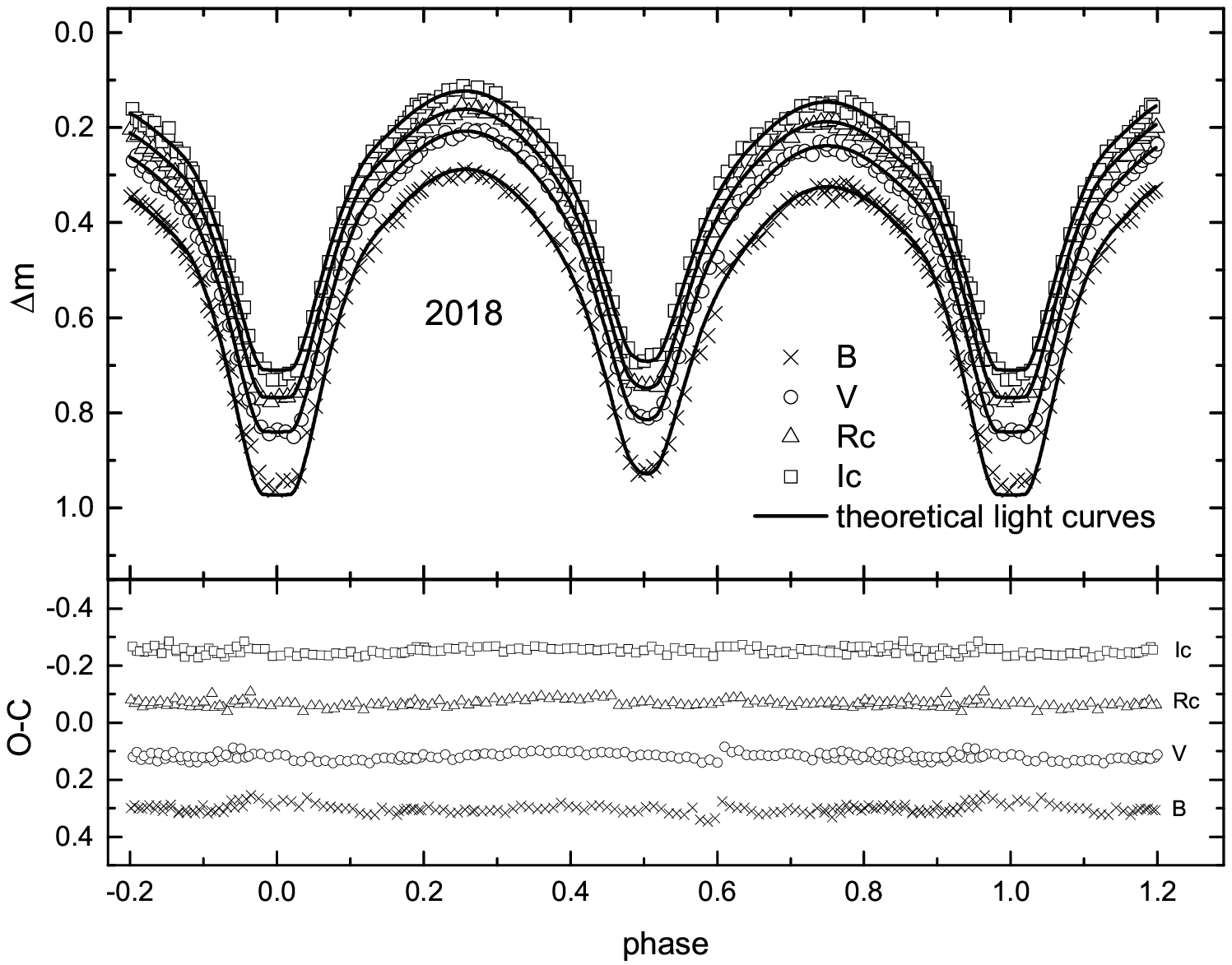}%
\includegraphics[width=0.5\textwidth]{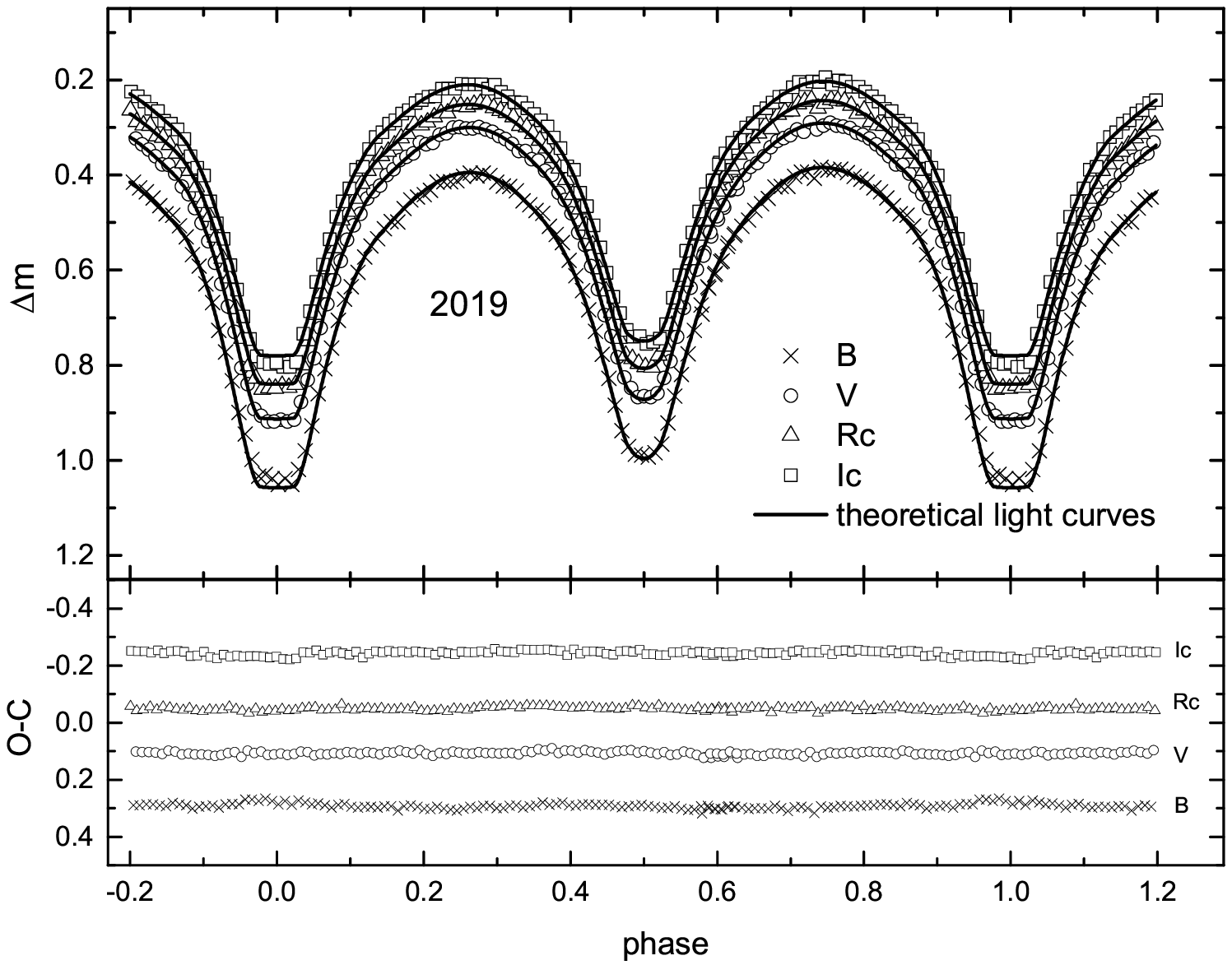}

\caption{The comparison between the synthetic and observed light curves for V342 UMa.}
\label{Fig5}
\end{center}
\end{figure}

\begin{figure}
\begin{center}
\includegraphics[width=0.33\textwidth]{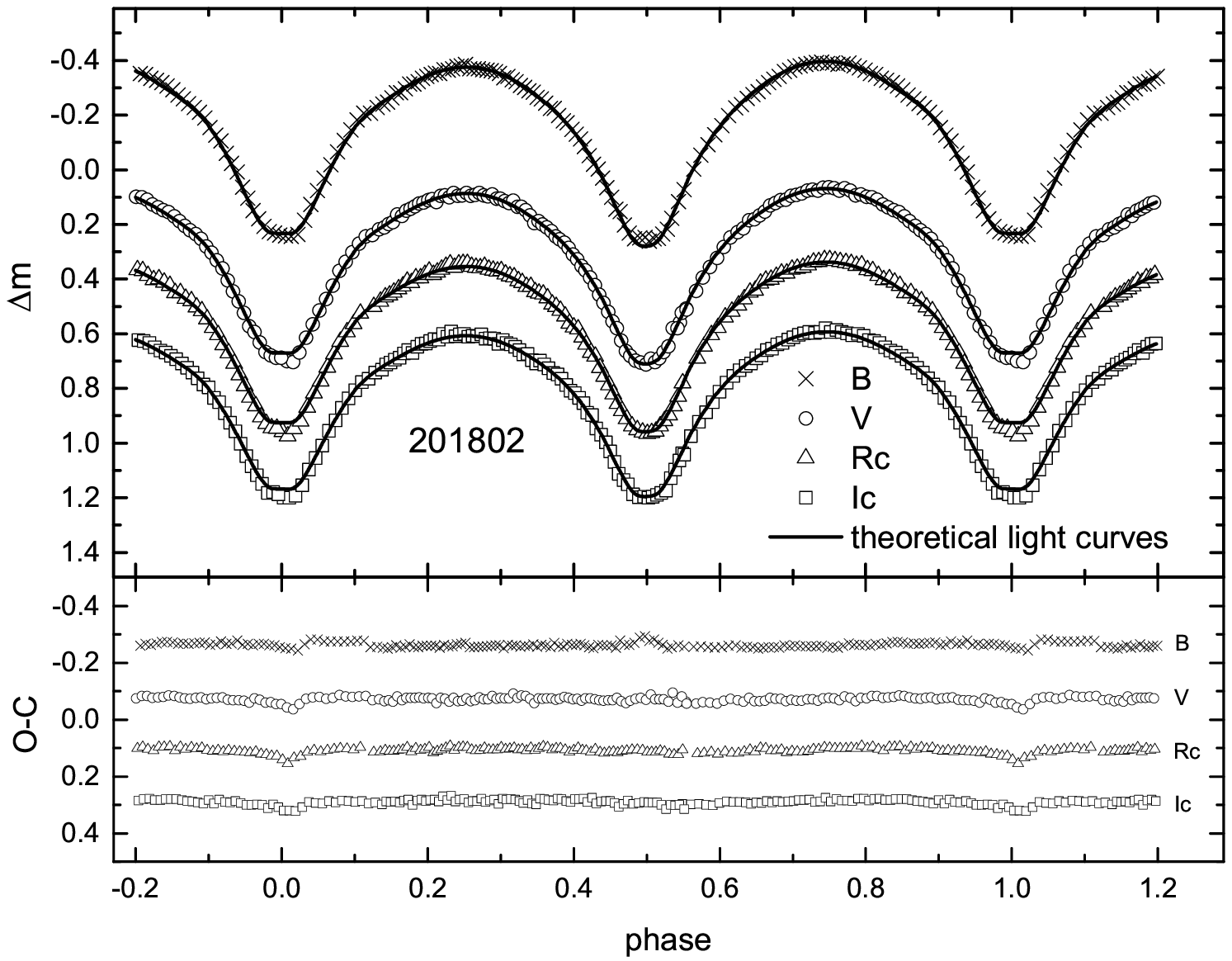}%
\includegraphics[width=0.33\textwidth]{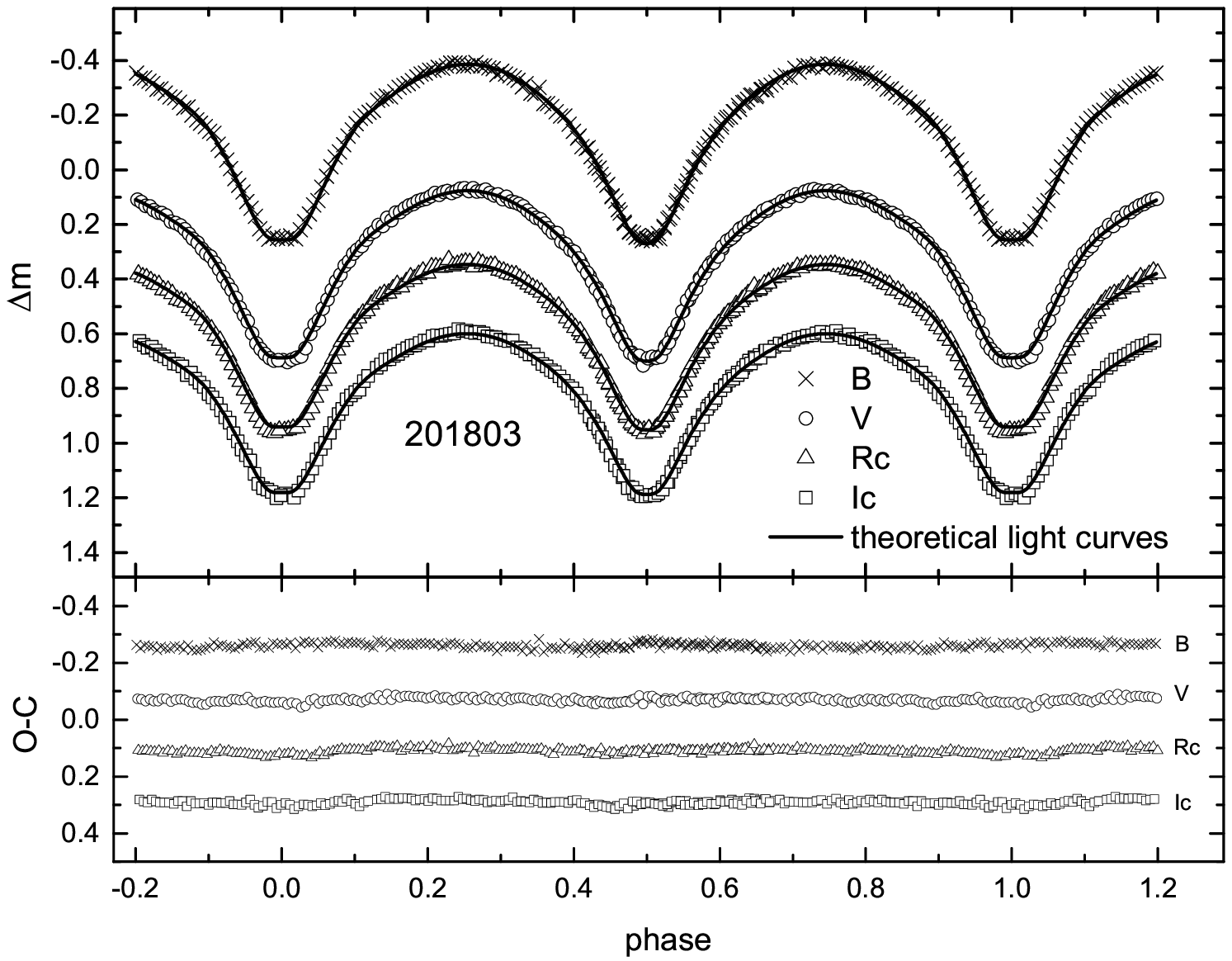}
\includegraphics[width=0.33\textwidth]{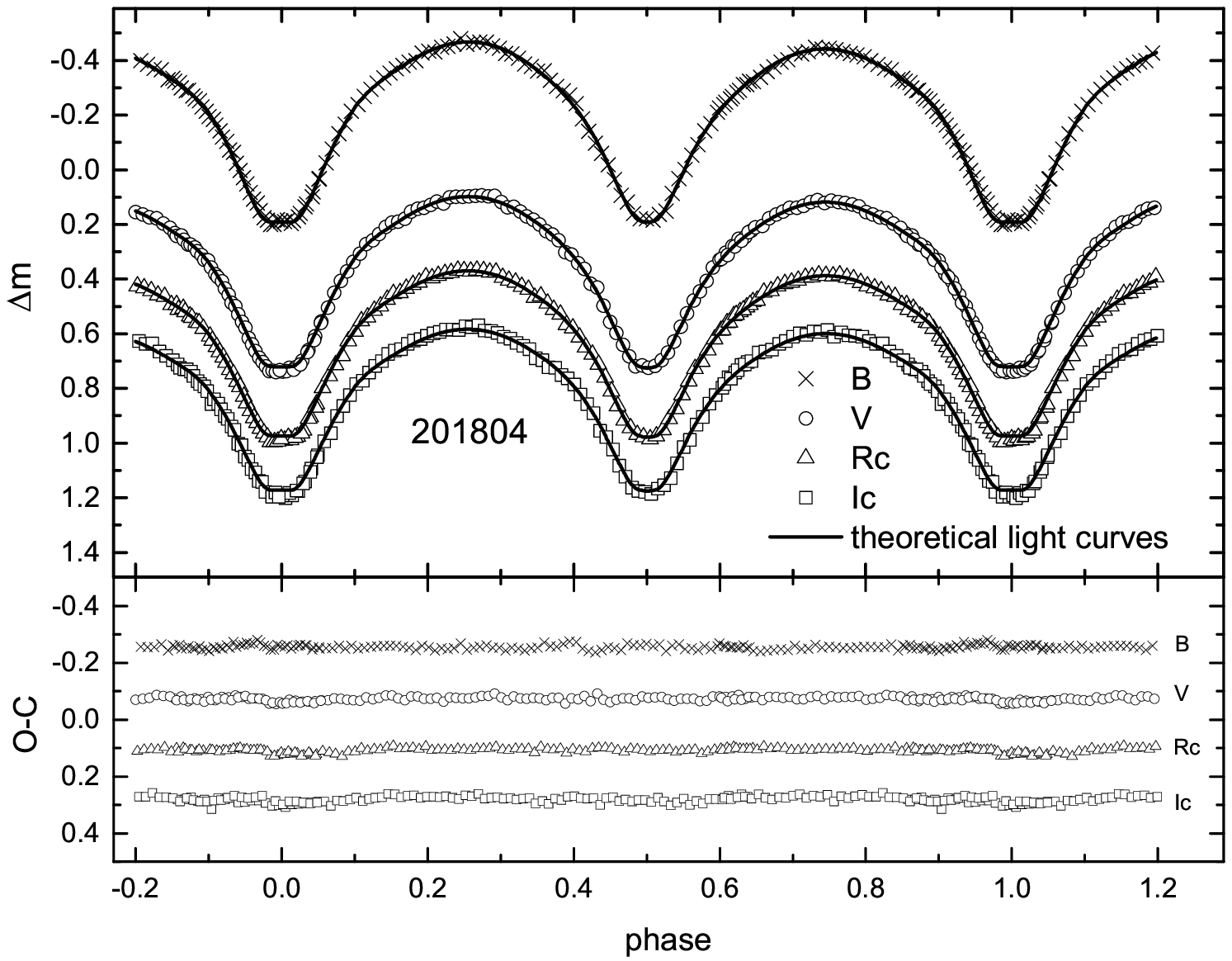}

\caption{The comparison between the synthetic and observed light curves for V509 Cam.}
\label{Fig6}
\end{center}
\end{figure}

Because both V342 UMa and V509 Cam were obtained two or more sets of complete light curves, and the light curves observed at different times are different. The physical parameters determined by different light curves may be different. Therefore, we chose one set of the complete light curves to determine the physical parameters, and the derived physical parameters were set as reference values to model the other light curves. For V342 UMa, the complete light curves observed in 2019 have higher quality comparing to those observed in 2018, so the 2019 light-curve was the chosen one. For V509 Cam, the complete light curves observed on March 06, 2018 are symmetric and have the highest precision among the three sets of light curves, so the 201803 light-curve was the chosen one. During the modeling, we used Mode 2 (detached configuration) for both targets at first and found that the solutions were quickly convergent at Mode 3 (contact configuration). The adjustable parameters are as follows: the orbital inclination, $i$, the temperature of the secondary, $T_2$, the dimensionless potential of the primary $\Omega_1$, and the monochromatic luminosity of the primary, $L_1$. Then, a series of of solutions with fixed values of mass ratio $q$ were carried out for them. The weighted sum of squared residuals, $\sum W_i(O-C)_i^2$, versus mass ratio $q$ of the two systems are respectively displayed in the left and right panels of Figure~\ref{Fig4}. As seen in Figure~\ref{Fig4}, a very sharp minimum was determined for V342 UMa at $q=2.8$, while that was derived for V509 Cam at $q=2.5$. These two values were set as initial values and adjustable parameters, new solutions were performed. When the solutions were convergent, the physical parameters were obtained. The light curves of V342 UMa is asymmetric, adding a cool spot on the less massive primary component can reproduce the asymmetric light curves. The derived physical parameters are listed in Table~\ref{Tab:photo-results}, and the synthetic light curves are respectively shown in Figure~\ref{Fig5} and Figure~\ref{Fig6}.

\begin{figure}
\begin{center}
\includegraphics[angle=0,scale=0.7]{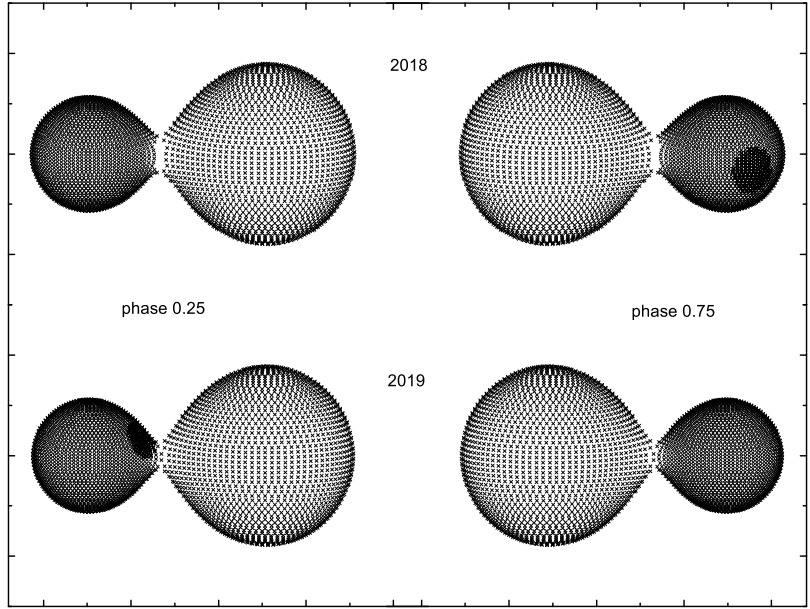}
\caption{Geometrical configurations of V342 UMa. }
\label{Fig7}
\end{center}
\end{figure}

\begin{figure}
\begin{center}
\includegraphics[angle=0,scale=0.7]{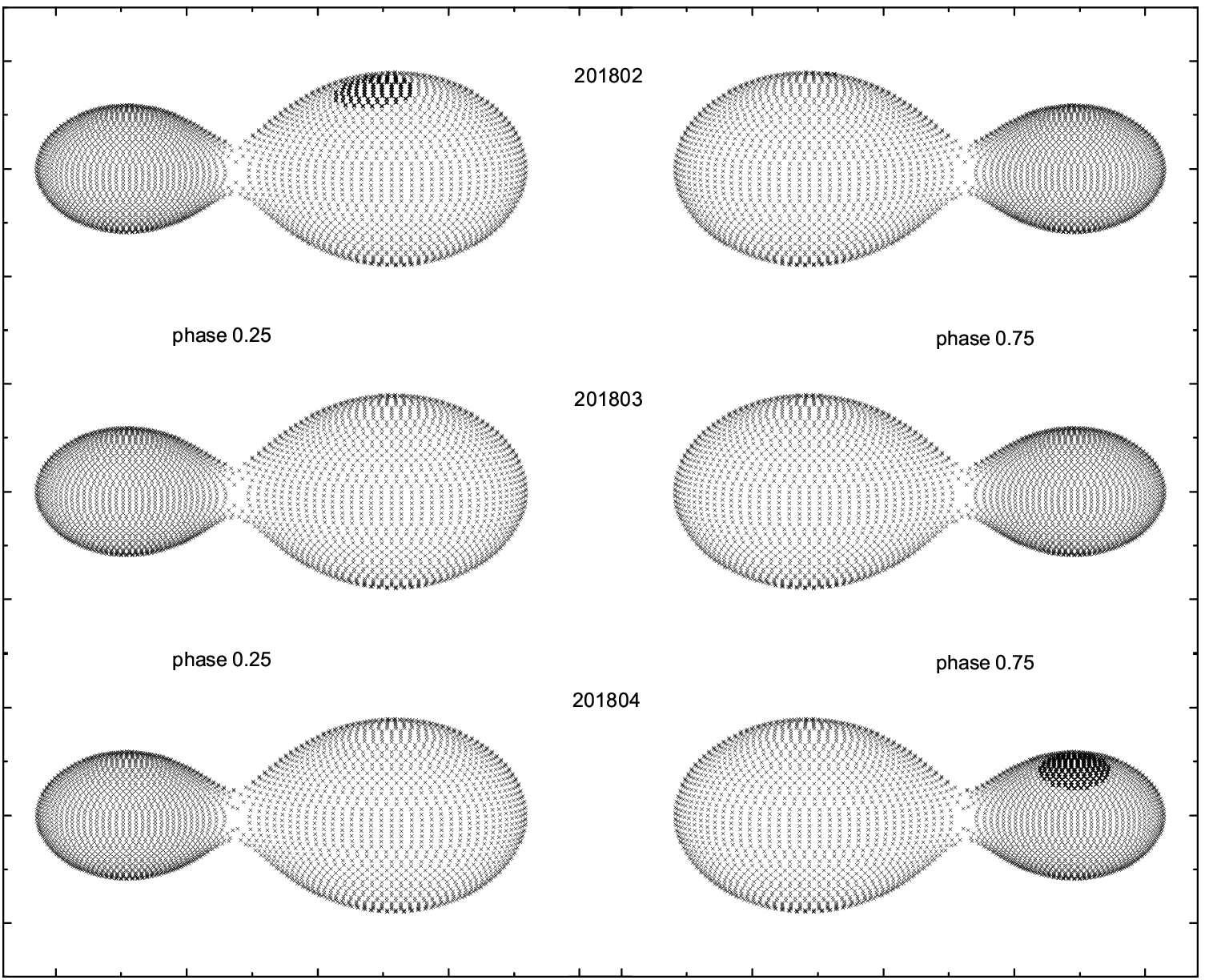}
\caption{Geometrical configurations of V509 Cam. }
\label{Fig8}
\end{center}
\end{figure}

To model the other light curves, the physical parameters derived above were set as reference values and the mass ratio $q$ was fixed. Due to the asymmetric light curves, spots model was applied. The best fitting results are also listed in Table~\ref{Tab:photo-results}, and the corresponding fitting curves are respectively displayed in Figure~\ref{Fig5} and Figure~\ref{Fig6}. According to the previous discussion, the physical parameters determined by the 2019 light-curve of V342 UMa and the 201803 light-curve of V509 Cam should be more reliable. Therefore, the physical parameters determined by the 2019 light-curve of V342 UMa and the 201803 light-curve of V509 Cam were adopted as the final results. The geometric configurations of the two systems are respectively plotted in Figure~\ref{Fig7} and Figure~\ref{Fig8}, the changes of the spot distributions can be clearly clarified.

\begin{table}
\small
\begin{center}
\caption{Photometric results of V342 UMa and V509 Cam}
\label{Tab:photo-results}
\begin{tabular}{l|cc|ccc}
\hline
Star &     \multicolumn{ 2}{c}{V342 UMa}  &    \multicolumn{ 3}{c}{V509 Cam} \\\hline

Parameters &     201901  &     201805 &    201803 &201802 &201804 \\\hline

$T_1$(K) &     $5902\pm6$ &     $5899\pm8$ &     $6523\pm4$ &     $6498\pm5$ &     $6560\pm5$ \\

$T_2$(K) &     $5662\pm11$ &     $5663\pm16$ &     $6433\pm9$ &     $6446\pm12$ &     $6415\pm11$ \\

$q$ & $2.748\pm0.015$ & 2.748(fixed) & $2.549\pm0.016$& 2.549(fixed) & 2.549(fixed)  \\

$i(^\circ)$ & $84.2\pm0.2$ & $82.9\pm0.3$ & $82.1\pm0.1$ & $82.3\pm0.1$ & $82.8\pm0.2$  \\

$\overline {L_2/L_1}$ & $2.040\pm0.113$ & $2.024\pm0.110$ & $2.145\pm0.037$ & $2.202\pm0.023$ & $2.073\pm0.058$ \\

$\Omega_1=\Omega_2$ & $6.219\pm0.019$ & $6.164\pm0.006$ & $5.816\pm0.021$ & $5.821\pm0.005$ & $5.828\pm0.005$ \\

 $r_1$ & $0.299\pm0.001$ & $0.305\pm0.001$ & $0.321\pm0.001$ & $0.319\pm0.001$ & $0.320\pm0.001$ \\

 $r_2$ & $0.476\pm0.002$ & $0.480\pm0.001$ & $0.481\pm0.003$ & $0.481\pm0.001$ & $0.481\pm0.001$ \\

 $f$ &  $10.0\pm3.1\%$ &  $19.0\pm1.0\%$ &  $32.1\pm3.4\%$ &  $31.3\pm0.9\%$ &  $30.1\pm0.9\%$   \\\hline

Spot&  on star 1 & on star 1&$-$  & on star 2 & on star 1 \\\hline

$\theta$(radian) & $1.273\pm0.092$ & $1.759\pm0.212$ & $-$ & $0.467\pm0.107$ &$0.583\pm0.132$ \\

$\phi$(radian)   & $5.896\pm0.064$ & $2.134\pm0.056$ & $-$ & $1.225\pm0.057$ &$1.658\pm0.039$ \\

 $r$(radian)     & $0.258\pm0.069$ & $0.358\pm0.085$ & $-$ & $0.298\pm0.064$ &$0.423\pm0.086$ \\

$T_f(T_d/T_0)$   & $0.832\pm0.099$ & $0.789\pm0.124$ & $-$ & $0.811\pm0.089$ &$0.824\pm0.078$ \\

\hline
\end{tabular}
\end{center}
\end{table}

\section{Discussion and conclusions}
\label{sect:discussion}

Two sets of complete $BVR_cI_c$ light curves of V789 Her and three sets of complete $BVR_cI_c$ light curves of V509 Cam were obtained and analyzed. We discovered that both of the two systems are W-subtype contact binaries, $q=2.748\pm0.015$ for V342 UMa and $q=2.549\pm0.016$ for V509 Cam. V342 UMa is a shallow contact binary with a fill-out factor of $f=10.0\pm3.1\%$, and V509 Cam is a medium contact binary with a contact degree of $f=32.1\pm3.4\%$. The two systems show totally eclipsing primary minima, the inclinations of them are higher than $82^\circ$, and the $q$-search curves exhibit very clear sharpness around the bottom. All these indicate that the photometric solutions derived only by the photometric light curves are reliable (e.g., \citealt{Pribulla+etal+2003, Terrell+Wilson+2005, Zhang+etal+2017}). Our photometric results of V342 UMa, such as the less massive component is the hotter one, and the reciprocal of our mass ratio is $1/q\sim0.364$,  are similar with those determined by \cite{Nelson+etal+2004}.
By collecting all available eclipsing times of V342 UMa and V509 Cam, we studied the period changes and obtained that the period of V342 UMa is secular decrease at a rate of $-1.02(\pm0.54)\times10^{-7}$ days/year and the period of V509 Cam is continuously increase at a rate of $3.96(\pm0.90)\times10^{-8}$ days/year.

\subsection{Absolute Parameters Estimation}
Because of the lack of radial velocity curves of V342 UMa and V509 Cam, we can not directly determine the absolute parameters. However, due to the Gaia mission (\citealt{Gaia Collaboration+etal+2018}), we can estimate the absolute parameters based on the distance. Firstly, the absolute magnitude of the two systems can be derived according to the parallax determined by Gaia mission and the relation $M_{V}=m_{V}-5\log D+5-A_V$, where $m_{V}$ is the $V$ band visual magnitude which can be derived from \cite{Gettel+etal+2006}, $D$ represent the distance which can be computed by the Gaia parallax, and $A_V$ is extinction value which can be derived from \cite{Chen+etal+2018}. Secondly, using the equations $M_{bol}=-2.5logL/L_\odot+4.74$ and $M_{bol}=M_V+BC_V$ ($M_{bol}$ is the absolute bolometric magnitude and $BC_V$ is the bolometric correction which can be interpolate from Table 5 of \cite{Pecaut+Mamajek+2013}, the total luminosity of the binary can be calculated. Thirdly, the luminosity of each component ($L_1$ and $L_2$) can be determined by the luminosity ratio $L_2/L_1$ listed in Table~\ref{Tab:photo-results}. Fourthly, assuming black-body emission($L=4\pi\sigma T^4R^2$), the radius of each component can be estimated, hence the semi-major axis $a$ can be obtained  based the absolute and relative radius of each component (an average value of $a$ was adopted). Finally, the mass of each component can be calculated by using Kepler's third law $M_1+M_2=0.0134a^3/P^2$ and the mass ratio $q$. Following these steps, we obtained the absolute parameters of V342 UMa and V509 Cam. The absolute parameters, along with the parameters needed in the calculation process, are listed in Table~\ref{Tab:abs-parameters}. This method provide an opportunity to estimate absolute parameters of contact binaries without radial velocity curve observations and can be applied to other contact binaries with reliable photometric solutions.

\begin{table}
\tiny
\begin{center}
\caption{Absolute parameters of V342 UMa and V509 Cam}
\label{Tab:abs-parameters}
\begin{tabular}{lcccccccccccc}
\hline
Parameters& $D$     &$V_{max}$    &$M_V$   &$BC_V$  &$M_{bol}$    &$L_1$       &$L_2$       &$R_1$       &$R_2$       &$a$         &$M_1$        &$M_2$       \\\hline
          & (pc)    & (mag)       &(mag)   &(mag)   &(mag)       &($L_\odot$) &($L_\odot$) &($R_\odot$) &($R_\odot$) &($R_\odot$) &($M_\odot$)  &($M_\odot$) \\\hline
V342 UMa  &$685.6$  &$13.418$     &$4.134$ &$-0.115$  &$4.019$      &$0.640$     &$1.303$     &$0.766$     &$1.188$     &$2.530$     &$0.490$      &$1.293$   \\
     &$\pm14.0$&$-$  &$\pm0.044$&$-$  &$\pm0.044$ &$\pm0.090$   &$\pm0.157$   &$\pm0.054$   &$\pm0.072$   &$\pm0.175$ &$\pm0.103$    &$\pm0.291$   \\
V509 Cam  &$674.0$   &$12.933$ &$3.647$  &$-0.041$  &$3.606$ &$0.904$   &$1.938$   &$0.746$   &$1.122$   &$2.328$ &$0.388$    &$0.997$   \\
     &$\pm8.6$ &$-$  &$\pm0.028$&$-$  &$\pm0.028$ &$\pm0.068$   &$\pm0.134$   &$\pm0.028$   &$\pm0.039$   &$\pm0.093$ &$\pm0.048$    &$\pm0.129$   \\
\hline
\end{tabular}
\end{center}
\end{table}

\subsection{The Secular Period Changes}
The period of V342 UMa is secular decrease at a rate of $-1.02(\pm0.54)\times10^{-7}$ days/year. Usually, the long-term period decrease is produced by conservative mass transfer or angular momentum loss (AML). If it is caused by conservative mass transfer, the mass transfer rate can be determined to be $dM_1/dt=3.01(\pm1.59)\times10^{-8}\,M_\odot$ yr$^{-1}$ by using the following equation,
\begin{eqnarray}
{\dot{P}\over P}=-3\dot{M_1}({1\over M_1}-{1\over M_2}) .
\end{eqnarray}
The positive sign indicates that the less massive primary component is receiving mass. Assuming the angular momentum and the total mass are constant and the more massive component transfers its mass on a thermal timescale, $\tau_{th}=3.39\times10^7$ years can be calculated using this relation $\tau_{th}={GM_2^2\over R_2L_2}$. Then, the mass transfer rate can be roughly derived to be $M_2/\tau_{th}=3.81\times10^{-8}\,M_\odot$ yr$^{-1}$, which is coincide with the result determined by Equation (6). Another possibility is the AML due to magnetic stellar winds. An approximate for calculating the period decrease rate was given by \cite{Guinan+Bradstreet+1988} as below,
\begin{eqnarray}
{dP\over dt}\approx -1.1\times10^{-8}q^{-1}(1+q)^2(M_1+M_2)^{-5/3}k^2\times(M_1R_1^4+M_2R_2^4)P^{-7/3},
\end{eqnarray}
where $k^2$ is the gyration constant. Taking $k^2=0.1$ from \cite{Webbink+1976} for low-mass main-sequence stars, we derived that the period decrease rate caused by AML is $-0.71\times10^{-7}$ days/year, which is similar to the observed value. Therefore, both the conservative mass transfer and AML can explain the long-term period decrease of V342 UMa. At present, we can not know which one is dominant only based on the observed period variation.

The period of V509 Cam is continuously increase at a rate of $3.96(\pm0.90)\times10^{-8}$ days/year. The long-term period increase is generally attributed to conservative mass transfer. Using Equation (6), we derived that the mass transfer rate is $dM_1/dt=-1.07(\pm0.24)\times10^{-8}\,M_\odot$ yr$^{-1}$. The negative sign exhibits that the less massive primary component is transferring mass to the more massive secondary one. V509 Cam is a late type contact binary, the AML due to magnetic stellar winds should also can happen in V509 Cam. Therefore, the determined mass transfer rate should be considered as a minimal value.

The secular period decrease rate, $-1.02(\pm0.54)\times10^{-7}$ days/year, of V342 UMa is very common in W-subtype contact binaries, such as $-1.69\times10^{-7}$ days/year for V502 Oph (\citealt{Zhou+etal+2016}), $-0.62\times10^{-7}$ days/year for GU Ori (\citealt{Zhou+etal+2018}), and $-1.78\times10^{-7}$ days/year for V1007 Cas (\citealt{Li+etal+2018}). With decreasing period, V342 UMa will evolve from the present shallow contact configuration to high fill-out contact state. The long-term period increase rate, $3.96(\pm0.90)\times10^{-8}$ days/year, of V509 Cam is also very common in W-subtype contact binaries, such as $5.09\times10^{-8}$ days/year for EP And (\citealt{Lee+etal+2013}), $7.7\times10^{-8}$ days/year for UX Eri (\citealt{Qian+etal+2007b}), and $6.5\times10^{-8}$ days/year for LR Cam (\citealt{Yang+Dai+2010}). With increasing period, V509 Cam may evolve into a broken-contact binary. There is another possibility that the long-term period variations of the two systems are only part of very long period periodic variations as a result of distant third body (\citealt{Liao+Qian+2010}), continuous observations of the two targets are required to confirm this in the future.

\subsection{Light Curve Variations of the Two Targets}
As seen in Figure~\ref{Fig1} and Figure~\ref{Fig2}, we can discover that the light curves of V342 UMa and V509 Cam display very clear variations. For V342 UMa, the light curves observed in 2018 show a positive  O'Connell effect (means the first light maximum, Max. I, is brighter than the second one, Max. II), while the light curves observed in 2019 reversed to exhibit a negative O'Connell effect (means Max. I is fainter than the Max. II). For V509 Cam, the light curves observed in February, 2018 show a negative O'Connell effect, then light curves observed in March, 2018 are almost symmetric. However, the light curves reverse to exhibit a positive O'Connell effect in April, 2018. The differences between Min.I and Min.II, Max.II and Max.I, Min.I and Max.I, and Min.II and Max.I were calculated and are listed in Table~\ref{Tab:differences}. Such changes in the light curves are very common in contact binaries, such as BX Peg (\citealt{Lee+etal+2004, Lee+etal+2009}), HH UMa (\citealt{Wang+etal+2015}), RT LMi (\citealt{Qian+etal+2008}). They are generally caused by magnetic activities and can be interpreted by the presence of spots. The changes of the light curves of V342 UMa and V509 Cam are both explained by spot variation.

\begin{table}
\footnotesize
\begin{center}
\caption{Light diffferences for V342 UMa and V509 Cam at different epochs}
\label{Tab:differences}
\begin{tabular}{lccccc}
\hline
Epoch	      &Filter	&Min.I - Min.II&	Max.II - Max.I&	Min.I - Max.I&	Min.II - Max.I\\ \hline
V342 UMa	  &     &         &        &        &                                     \\ \hline
2018 May    &$B$	 &0.041 	  &0.034 	  &0.673 	&0.632                              \\
      &$V$	 &0.027 	  &0.031 	  &0.631 	&0.604                              \\
      &$R_c$	 &0.020 	  &0.027 	  &0.607 	&0.587                              \\
      &$I_c$	 &0.038 	  &0.026 	  &0.609 	&0.571                              \\
2019 Jan    &$B$	 &0.055 	  &-0.012 	&0.655 	&0.600                              \\
	      &$V$	 &0.043 	  &-0.009 	&0.614 	&0.572                              \\
	      &$R_c$	 &0.036 	  &-0.008 	&0.591 	&0.555                              \\
	      &$I_c$	 &0.045 	  &-0.007 	&0.584 	&0.539                              \\\hline
V509 Cam    &     &         &        &        &                                     \\\hline
2018 Feb    &$B$	 &-0.008 	&-0.021 	&0.620 	&0.628                                \\
	      &$V$	 &-0.007 	&-0.018 	&0.611 	&0.618                                \\
	      &$R_c$	 &0.000  	&-0.016 	&0.605 	&0.605                                \\
	      &$I_c$	 &0.004  	&-0.015 	&0.593 	&0.588 	                              \\
2018 Mar    &$B$	 &-0.004 	&0.001 	&0.642 	&0.646                                  \\
	      &$V$	 &-0.009 	&-0.001 	&0.610 	&0.619                                \\
	      &$R_c$	 &0.003  	&0.000 	&0.608 	&0.605                                  \\
	      &$I_c$	 &0.003  	&0.000 	&0.591 	&0.588                                  \\
2018 Apr    &$B$	 &0.007  	&0.026 	&0.660 	&0.653                                  \\
	      &$V$	 &0.013  	&0.021 	&0.641 	&0.628                                  \\
	      &$R_c$	 &0.012  	&0.018 	&0.621 	&0.609                                  \\
	      &$I_c$	 &0.021  	&0.016 	&0.613 	&0.592                                  \\
\hline
\end{tabular}
\end{center}
Note. Min.I, Min.II, Max.I, and Max.II respectively denote the primary light minimum, the secondary light minimum, the light maximum after Min. I, and the light maximum after Min. II.
\end{table}

\subsection{The Evolutionary Status}

\begin{figure}
\begin{center}
\includegraphics[angle=0,scale=0.5]{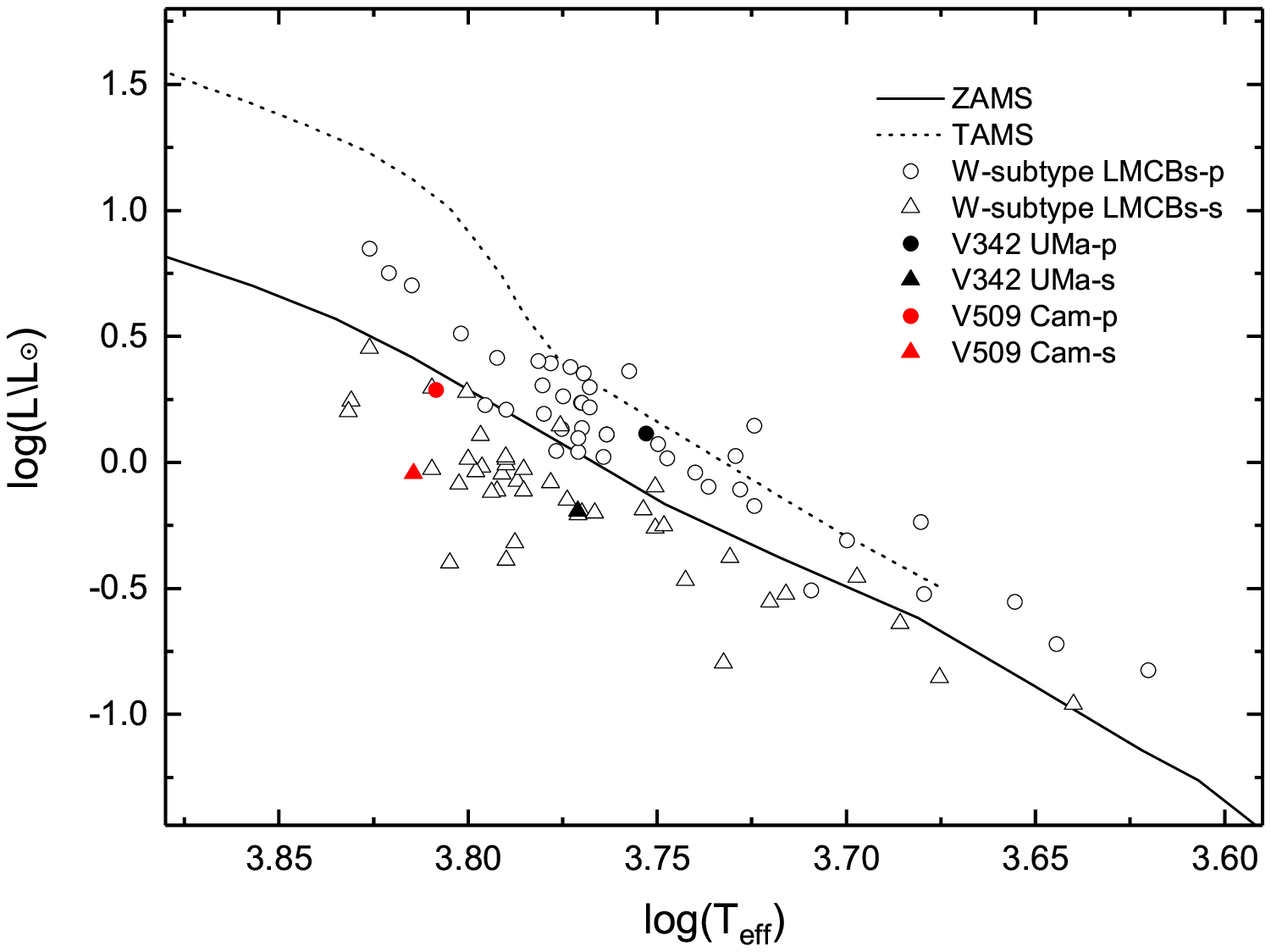}
\includegraphics[angle=0,scale=0.45]{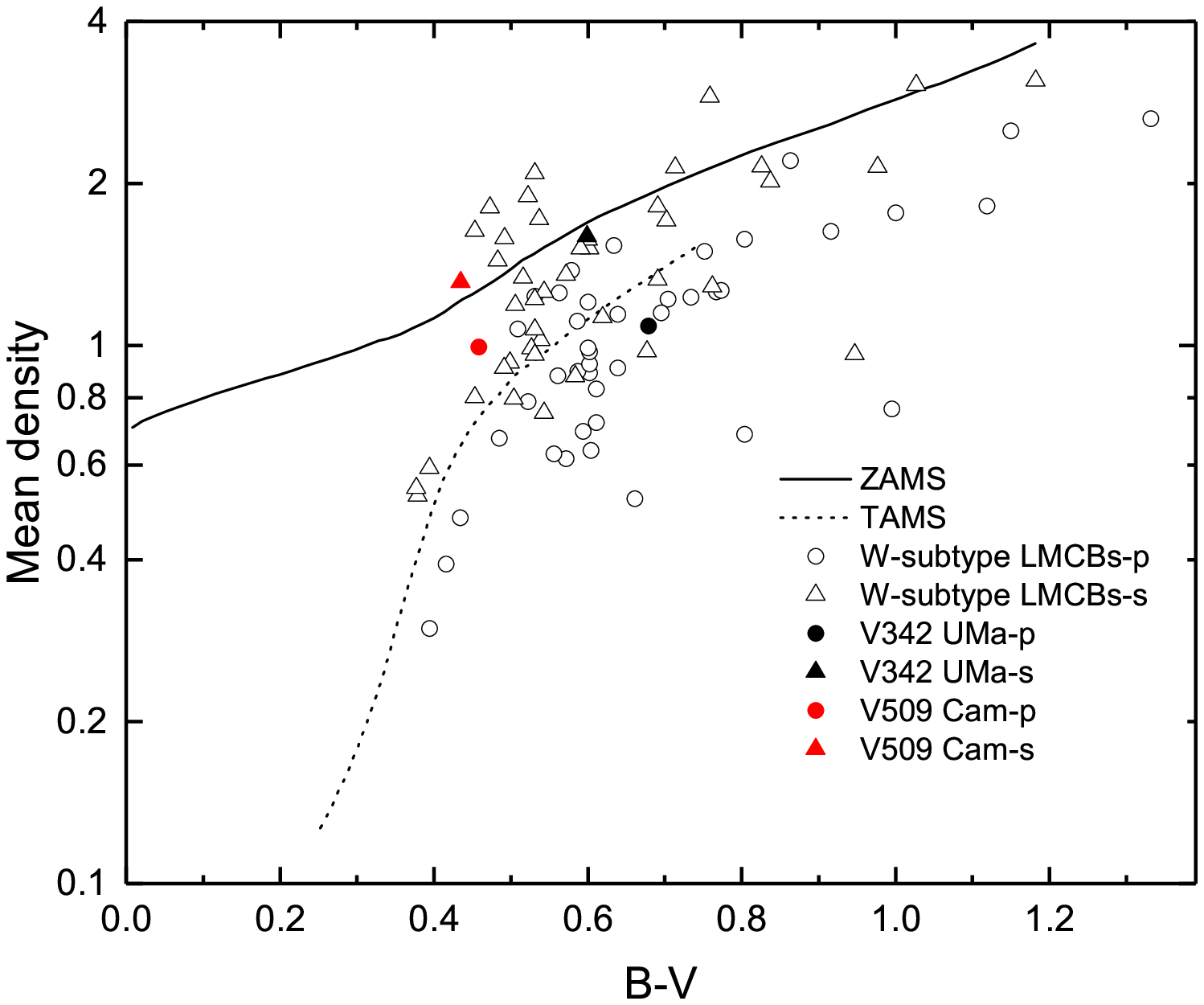}
\caption{The H-R diagram and C-D diagram. The solid and dotted lines display the ZAMS and TAMS, respectively. The circles refer to the more massive components (p), while the triangles represent the less massive ones (s).}
\label{Fig9}
\end{center}
\end{figure}

To study the evolutionary status of the two stars, the Hertzsprung-Russell (H-R) diagram and the color-density (C-D) diagram were constructed and are shown in the left and right panels of Figure~\ref{Fig9}, respectively. The zero age main sequence (ZAMS) and the terminal age main sequence (TAMS) in the H-R diagram were taken from \cite{Girardi+etal+2000}, while the ZAMS and TAMS in the C-D diagram were come from \cite{Mochnacki+1981}. In order to compare with other W-subtype contact binaries, the W-subtype low mass contact binaries (LMCBs) listed in \cite{Yakut+Eggleton+2005} are also displayed in Figure~\ref{Fig9}. The horizontal ordinate of the right panel of Figure~\ref{Fig9} is color-index, we converted the temperatures of the components of all systems including V342 UMa and V509 Cam to color-index based on Table 5 of \cite{Pecaut+Mamajek+2013}. The mean densities of the components were calculated using the relation provided by \cite{Mochnacki+1981},
\begin{eqnarray}
\overline{\rho_1}={0.079\over V_1(1+q)P^2}\,g\,cm^{-3}, \qquad \overline{\rho_2}={0.079q\over V_2(1+q)P^2}\,g\,cm^{-3}.
\end{eqnarray}
In Figure~\ref{Fig9}, the ZAMS and TAMS are labeled as solid and dotted lines, respectively, and the circles refer to the more massive components (p), while the triangles represent the less massive ones (s). Both the H-R diagram and the C-D diagram reveal that the components of V342 UMa and V509 Cam are consistent with those of other W-subtype contact systems. The less massive components are close to the ZAMS, meaning that they are main-sequence or little evolved stars, while the more massive ones are located near the TAMS, indicating that they are at an advanced evolutionary stage. The evolutionary status of the components of W-subtype contact systems is similar with that of the A-subtype ones. Therefore, W-subtype phenomenon is still an open question. More observations and investigations on the two subtype contact binaries are needed.

In this paper, the investigations of the light curves and period variations of V342 UMa and V509 Cam are presented. We found that both of the two systems are W-subtype contact binaries and show very strong light curve changes. The period changes analysis reveal that V342 UMa shows long-term period decrease which can be caused by conservative mass transfer or AML due to magnetic stellar winds and that V509 Cam exhibits long-term period increase which can be attributed to conservative mass transfer. The absolute parameters of the two binaries were obtained based on the Gaia distances. In order to identify cyclic period changes of them, further observations are required.

\begin{acknowledgements}
This work is supported by Chinese Natural Science Foundation (No. 11703016), and by the Joint Research Fund in Astronomy (No. U1431105) under cooperative agreement between the National Natural Science Foundation of China (NSFC) and Chinese Academy of Sciences (CAS), and by the program of the Light in China¡¯s Western Region, No. 2015-XBQN-A-02, and by the Natural Science Foundation of Shandong Province (Nos. ZR2014AQ019, JQ201702), and by Young Scholars Program of Shandong University, Weihai (Nos. 20820162003, 20820171006), and by the program of Tianshan Youth (No. 2017Q091), and by the Open Research Program of Key Laboratory for the Structure and Evolution of Celestial Objects (No. OP201704). Thanks the referee very much for the very helpful comments and suggestions to improve our manuscript.

We acknowledge the support of the staff of the Xinglong 85cm
telescope, NOWT, WHOT and NEXT. This work was partially supported by the Open Project Program of the Key
Laboratory of Optical Astronomy, National Astronomical Observatories, Chinese Academy of Sciences.

This paper makes use of data from the DR1 of the WASP data (\citealt{Butters+etal+2010}) as provided by the WASP consortium,
and the computing and storage facilities at the CERIT Scientific Cloud, reg. no. CZ.1.05/3.2.00/08.0144
which is operated by Masaryk University, Czech Republic.

This work has made use of data from the European Space Agency (ESA) mission
{\it Gaia} (\url{https://www.cosmos.esa.int/gaia}), processed by the {\it Gaia}
Data Processing and Analysis Consortium (DPAC,
\url{https://www.cosmos.esa.int/web/gaia/dpac/consortium}). Funding for the DPAC
has been provided by national institutions, in particular the institutions
participating in the {\it Gaia} Multilateral Agreement.
\end{acknowledgements}

%
%
%


\label{lastpage}

\end{document}